\documentclass[twocolumn]{emulateapj}
\usepackage{graphicx}
\usepackage{amssymb}
\usepackage{epstopdf}
\usepackage{ctable}
\newcommand{\degree}{\mbox{$^{\circ}$}}
\usepackage{multirow}
\usepackage{longtable}
\usepackage{amsmath}
\usepackage{multirow}
\usepackage{threeparttable}
\usepackage{threeparttablex}
\usepackage{environ}
\usepackage{natbib}
\usepackage{pslatex}
\shorttitle{Decoupled Clouds in Stripped Virgo Spirals}
\shortauthors{Abramson \& Kenney}

\begin{document}

\title{HST Imaging of Decoupled Dust Clouds in the Ram Pressure Stripped Virgo Spirals NGC 4402 and NGC 4522}
\author{Anne Abramson \& Jeffrey D. P. Kenney}
\affil{Department of Astronomy, Yale University, P.O. Box 208101, New Haven, CT 06520, U.S.A.}
\email{anne.abramson@yale.edu, jeff.kenney@yale.edu}

\begin{abstract}
We present the highest-resolution study to date of the ISM in galaxies undergoing ram pressure stripping, using HST BVI imaging of NGC 4522 and NGC 4402, Virgo Cluster spirals that are well-known to be experiencing ICM ram pressure.  We find that throughout most of both galaxies, the main dust lane has a fairly well-defined edge, with a population of GMC-sized (tens- to hundreds-of-pc scale), isolated, highly extincting dust clouds located up to $\sim$1.5 kpc radially beyond it.  Outside of these dense clouds, the area has little or no diffuse dust extinction, indicating that the clouds have decoupled from the lower-density ISM material that has already been stripped.  Several of the dust clouds have elongated morphologies that indicate active ram pressure, including two large (kpc-scale) filaments in NGC 4402 that are elongated in the projected ICM wind direction.  We calculate a lower limit on the HI + H$_2$ masses of these clouds based on their dust extinctions and find that a correction factor of $\sim$10 gives cloud masses consistent with those measured in CO for clouds of similar diameters, probably due to the complicating factors of foreground light, cloud substructure, and resolution limitations.  Assuming that the clouds' actual masses are consistent with those of GMCs of similar diameters ($\sim 10^4-10^5$ M$_{\odot}$), we estimate that only a small fraction ($\sim$1-10\%) of the original HI + H$_2$ remains in the parts of the disks with decoupled clouds.  Based on H$\alpha$ images, a similar fraction of star formation persists in these regions, 2-3\% of the estimated pre-stripping star formation rate.  We find that the decoupled cloud lifetimes may be up to 150-200 Myr. 

\end{abstract}


\section{Introduction}

Numerous observations of clusters and individual galaxies indicate that ram pressure stripping is an effective mechanism for galaxy transformation with profound effects on a galaxy's total gas content \citep[e.g.,][]{chung09, vollmer12} interstellar medium (ISM) spatial distribution \citep[e.g.,][]{kenney04}, stellar populations \citep[e.g.,][]{crowl08}, and star formation activity and spatial distribution \citep[e.g.,][]{abramson11,vollmer12,merluzzi13}.  In rich clusters like Coma, ram pressure is capable of transforming spirals into gas-deficient, quiescent disk galaxies - some Coma Cluster spirals have been almost completely ram pressure stripped \citep{smith10,yagi10}, with their stripped ISM forming long tails extending from undisturbed stellar disks.  The Virgo cluster has a more diffuse intracluster medium (ICM) but has a number of spirals with long HI tails \citep{chung07} and radially truncated gas and star formation inside intact stellar disks \citep{koopmann04a}.  The ram pressure in Virgo is capable of partially stripping larger spirals and completely stripping dwarf galaxies \citep{kenney13}.  

There is no shortage of observations of how ram pressure stripping affects the ISM on $\geq$ kpc spatial scales, but sensitivity and resolution have limited our ability to observe gas on giant molecular cloud (GMC) scales (10-100 pc) in galaxies undergoing stripping.  In this paper, we use the Hubble Space Telescope (HST) to study GMC-scale dust clouds in the nearest spiral galaxies undergoing significant ram pressure stripping, which are located in the Virgo cluster.  High-resolution observations will reveal how ram pressure affects the complex, multi-phase ISM.  Relatively sharp radial cutoffs are observed in the H$\alpha$ emission of many cluster galaxies, leading to the concept of a ``gas truncation radius" \citep{koopmann04a} beyond which the disk is virtually devoid of the ISM necessary to sustain star formation.  However, the transition between unstripped inner disk and stripped outer disk is not always perfectly sharp.  One of the galaxies in this work, NGC 4402, has two dramatic dust filaments (0.5 kpc and 1 kpc long) outside the main gas truncation boundary that appear to have remained in the disk after the surrounding ISM was stripped and to have been elongated in the ICM wind direction \citep{crowl05}.  The existence of these clouds is consistent with the known physics of ram pressure stripping, which predicts that denser clouds at a given radius should remain in the disk as the less-dense ISM around them is stripped \citep[e.g.,][]{tonnesen09}.  

The ISM has a wealth of substructure on sub-kpc spatial scales, and there are almost no observations of how this substructure is affected by ram pressure stripping, other than the large dust filaments examined by \citet{crowl05}.  There is observational evidence that some of a galaxy's CO can decouple from the less-dense HI gas in the ISM  \citep[e.g.,][]{vollmer05,vollmer08}, but the resolution in these studies is not sufficient to detect individual clouds.  There are many unanswered questions about the smaller clouds that are closer in size to typical GMCs.  Under what conditions do these clouds decouple from the rest of the ISM during stripping?  What fraction of the ISM mass decouples, and how long do the decoupled clouds survive?  What are the ranges of diameter and mass for the decoupled clouds?  Do clouds show evidence of ram pressure acting on them - for instance, are the clouds elongated in the direction of the ICM wind?  

Understanding the effect of ram pressure stripping on molecular gas and star formation is crucial if we are to determine the role ram pressure stripping plays in galaxy evolution.  Galaxies tend to have a bimodal distribution in color-magnitude diagrams \citep[e.g.,][]{strateva01, blanton03,bell04} - early-type galaxies with less star formation and older stellar populations fall on the ``red sequence," and late-type galaxies with active star formation tend to be in the ``blue cloud."  In between these distributions is the ``green valley," a sparsely populated region thought to be occupied by galaxies whose star formation is in the process of being quenched.  \citet{hogg04} show the galaxy color-magnitude diagram as a function of environment and find that in denser environments, there are fewer blue galaxies and more red and green galaxies, many of which have significant disks.  This is consistent with ram pressure stripping in clusters transforming star-forming blue disk galaxies into red disk galaxies, which have little or no star formation.  High-resolution observations may reveal where, and how efficiently, ram pressure quenches star formation.  In order to understand the effects of stripping on star formation, we must find out what happens to molecular clouds as the disk is stripped.  Since some dense clouds may be left behind after an initial stripping episode \citep[e.g.,][]{crowl05}, active star formation may still be possible in disks that appear stripped in lower-resolution observations.  Is there any star formation associated with clouds that remain in the disk after the surrounding ISM has been stripped away?  Is the star formation rate enhanced in gas exposed to active ram pressure?  Or does star formation cease all at once when the less-dense ISM is stripped at a given radius?

Observations of dust extinction using HST will allow us to answer the above questions in greater detail than ever before.  Dust extinction is a very sensitive tracer of low-column-density clouds, and HST has the high spatial resolution necessary to detect relatively small clouds - at the distance of the Virgo cluster, we can detect clouds down to $\sim$10 pc diameter and $\sim$$10^4 M_{\odot}$.  Dust extinction does have limitations as a tracer of gas mass - the extinction-to-mass conversion has various uncertainties \citep[e.g.,][]{thompson04}, and we can only detect clouds located between us and a significant fraction of the disk light.  However, dust extinction provides a reasonable lower limit for cloud masses, and HST observations of dust extinction far exceed the spatial resolution of of pre-ALMA CO observations.  


In this paper, we present the first detailed observations of GMC-sized dust clouds in the disks of galaxies undergoing ram pressure stripping.  In Section \ref{summaryofwork}, we describe the two Virgo Cluster galaxies used for this study, NGC 4402 and NGC 4522, which are both well-documented cases of ram pressure stripping.  In Section \ref{obs}, we describe the observations and processing of our HST optical broadband images.  In Section \ref{results}, we describe the identification of the dust clouds and the measurement of their sizes, masses, and locations within the galaxy.  In Section \ref{discussion}, we introduce the concept of a ``transition zone" where the outer disk has been stripped of all but the densest clouds.  We then estimate the fraction of the original ISM mass that remains in the transition zones of NGC 4522 and NGC 4402.  Using ground-based H$\alpha$ images, we estimate the current and pre-stripping star formation rates in the transition zones.  We constrain the survival times of the decoupled transition zone clouds and estimate average stripping rates for the two transition zones.  Finally, we discuss the nature and origin of extraplanar decoupled clouds.  In Section \ref{conclusions}, we summarize our findings.  A subsequent paper (hereafter Paper II) will examine the galaxies' large-scale dust complexes and radio continuum emission.  Throughout this paper, we assume a distance to the Virgo cluster of 16.5 Mpc \citep{mei07}.  

\section{Summary of Existing Work and Comparison of the Two Galaxies}
\label{summaryofwork}

The Virgo spirals NGC 4402 and NGC 4522 are two of the clearest and best-studied cases of ICM-ISM stripping in action.  
In many of their basic physical and observational properties, NGC 4402 and NGC 4522 are very similar galaxies (Table \ref{the2gals}).  Both are Virgo Cluster Sc galaxies with similar optical luminosities and HI masses (Table \ref{the2gals}).  They are both highly-inclined (80\degree\, and 78\degree, respectively) and strongly HI-deficient (as determined by comparing their HI masses to that of an average isolated galaxy of the same optical diameter using the formula of \citealt{haynes84}).  Both display strong signatures of active ram pressure stripping: HI distributions which are truncated well inside undisturbed stellar disks, one-sided extraplanar tails of HI and radio continuum \citep{crowl05,kenney04,vollmer04,vollmer10}, and radially truncated star formation \citep{koopmann04a,crowl05}.  Evidence of ongoing ram pressure is also present in the form of ridges of polarized radio continuum emission, which probably indicate ISM compression on the leading sides of the ram pressure interactions \citep{vollmer04,vollmer07,vollmer08}.  Beyond the polarized ridges, the radio emission is weaker than would be expected based on the observed FIR distribution and the FIR-radio correlation in galaxies \citep{murphy09}.  These ``local radio deficit regions" are probably regions where the densities of halo cosmic ray electrons and magnetic fields are lower than in an undisturbed galaxy, due to ram pressure.

  The two galaxies differ in their stripping histories, viewing angles, and cluster locations.  Differing amounts of extraplanar material indicate that the galaxies have had somewhat different stripping histories.  NGC 4522 has much more extraplanar ISM and star formation, with 40\% of its HI gas and 10\% of its H$\alpha$ emission above the disk plane, versus 6\% and 0.3\% respectively in NGC 4402.  NGC 4522 also has a smaller HI truncation radius (0.4 $R_{25}$) than NGC 4402 (0.6-0.7 $R_{25}$). Stellar population analysis of the outer disks of the two galaxies shows that star formation in the outer disk of NGC 4522 was quenched 50 - 100 Myr ago \citep{crowl08}, wheres NGC 4402's outer disk was quenched $\sim$200 Myr ago.

  The two galaxies provide us with different viewing angles of the stripping process - NGC 4522 is moving away from us in the cluster, so we see the trailing side, and NGC 4402 is moving towards us in the cluster, so we see the leading side.  The galaxies' motion through the cluster is particularly relevant to our observations of optically thick ISM tracers, such as dust extinction, since we can only see dust in the disk plane on the near side.  The viewing angle and galaxy motion are further discussed in Section \ref{cloudloc}.

  In simple models, ram pressure is strongest near the cluster center, because the strength of ram pressure is proportional to the density of the ICM and the square of the velocity difference between the galaxy and ICM.  NGC 4402 is located in a central region of Virgo,1.4$\degree$ (0.4 Mpc) in projection from the giant elliptical M87 at the cluster center.  Strong X-ray emission at NGC 4402's location \citep{bohringer04} indicates hot, dense ICM capable of ram pressure stripping \citep[e.g.,][]{kenney08}.  NGC 4402 is also the closest spiral galaxy in projection to the giant elliptical M86, at a projected distance of only 43 kpc (9$\arcmin$), and the velocity difference of $\sim$450 km s$^{-1}$ between the two galaxies \citep{chung09,kenney08} make NGC 4402 a plausible member of the M86 subcluster.  NGC 4522, in contrast, is located 0.8 Mpc in projection from M87, where the estimated pressure of a smooth, static ICM would be a factor of 10 too low to cause the observed HI stripping \citep{kenney04}.  The young ages of the stellar population in the gas-stripped outer disk \citep{crowl08}, as well as the evidence of active ram pressure (for example, the radio deficit noted by \citet{murphy09}), indicate that the galaxy is being stripped locally at intermediate clustercentric radii, possibly by substructure within the ICM \citep{kenney04}, rather than having been stripped long ago during a passage near the cluster core.  

\begin{table}[h]
\begin{threeparttable}
\caption{Galaxy Properties.  }

\begin{tabular}{|c|c|c|c|}
\hline
\hline
Galaxy & NGC 4402 & NGC 4522  & References \\
\hline
inclination  &    80$\degree$     &  78$\degree$\  & a, b   \\
$M_B$ (mag)   &  -18.55    &     -18.11 &   c   \\
Type   &     Sc   &     SB(s)cd  &    c   \\
$D_{25}$ ($\arcmin$) [kpc] &  3.89 [18.6]  &      3.72 [17.8] &  c    \\
$D_{M87}$ (deg)  &  1.4     &     3.3  &    c   \\
$M_{HI}$ ( M$_{\odot}$)  &   3.7$\times10^8$    &      3.4$\times10^8$  &   e \\
$def_{HI}$  &  0.74    &       0.86   &  e  \\
$V_r$ (optical) (km s$^{-1}$)  &     236  &      2331  &   c  \\
HI Truncation Radius (units of $R_{25}$) &   0.6-0.7   &    0.4   & a, f \\
$V_{max} $  (optical) km s$^{-1}$ & 130 &  130 &  g  \\
$W^{50\%}_{HI}$ km s$^{-1}$ (uncorrected for $i$) & 249    &   214 &   e   \\
v$_{gal}$ - v$_{Virgo}$ (km s$^{-1})$  & -847  &    1250  &     \\
Extraplanar HI percentage  &  6\% &  40\% & a, f      \\
$M_{H_2}$ (M$_{\odot}$)&     2.8$\times10^9 $  &     3$\times10^8$ &  h, i   \\
$L_{H\alpha + N[II]}$ (erg s$^{-1}$)  &  $7.9\times10^{40}$  &  $1.3\times10^{40}$  &  a, j    \\
Extraplanar H$\alpha$ percentage   &  0.3\% &    10\%  &  a, f    \\
$S_{\nu}$ (20 cm) (Jy) & 0.067  &  0.027  &   k    \\

\hline
\end{tabular}
\begin{tablenotes}
\item Where two references are given, the first is for NGC 4402. \label{the2gals}
\item \textbf{References.} (a) \citet{crowl05}, (b) this work, (c) RC3, \citet{rc3}, (d) \citet{koopmann04b}, (e) \citet{chung09}, (f) \citet{kenney04}, (g) \citet{rubin99}, (h) \citet{kenney89}, (i) \citet{vollmer08}, (j) \citet{kenney99}, (k) \citet{murphy09}
\end{tablenotes}

\end{threeparttable}
\end{table}

\section{Observations and Data Reduction}
\label{obs}
\subsection{Observations and Processing}

     Observations of NGC 4522 were made on November 21-22, 2003 with
     the Advanced Camera for Surveys aboard Hubble Space Telescope.\footnote{Based on observations made with the NASA/ESA Hubble Space Telescope, obtained at the Space Telescope Science Institute, which is operated by the Association of Universities for Research in Astronomy, Inc., under NASA contract NAS 5-26555. These observations are associated with program $\#$9773 and $\#$10528.} 
     Images were taken in the F435W, F606W and F814W bands, with total
     exposure times of 7356 seconds, 2358 seconds, and 2268 seconds
     respectively.  Hereafter, we refer to these filters as B, V, and I.  The data consist of six images in a 3-point line
     dither pattern for the B band, and three images each in
     the same dither pattern for the V and I bands.  
     
     Observations of NGC 4402 were made on January 22, 2006 with
     the Advanced Camera for Surveys.  The galaxy was also observed in the F435W, F606W 
     and F814W bands, with total exposure times of 7678 seconds, 2337 seconds, and 2372 seconds.  
     The data comprises six images with a 3-point line
     dither pattern for the B band, and three images each in
     the same dither pattern for the V and I bands.  
     
     After standard pipeline processing and alignment of the images, sky
     levels were subtracted for each of the amplifiers separately, and
     weight maps were created.  Median images were produced for each
     band using the task IMCOMBINE before drizzling them together, to
     achieve better removal of cosmic rays from the gap between the
     two chips of ACS.  The separate images were then combined using
     the task MULTIDRIZZLE within IRAF/STSDAS v3.2. 
     
\subsection{Revealing Small-Scale Features in the Optical Image Using Unsharp Masking and Modeling}
 We tried both unsharp masking and model subtraction to remove the galaxy's large-scale luminosity gradients and highlight small-scale features such as dust clouds.  The techniques yielded similar results, but unsharp masking proved slightly better at enhancing the contrast between the local stellar disk and the individual dust clouds we examine in Section \ref{cloudident}.  Since the model subtraction process was more labor-intensive than unsharp masking but did not yield better results, we did not pursue it for NGC 4402.  However, model subtraction did yield more precise estimates of NGC 4522's exponential disk center, inclination, and position angle than currently exist in the literature.  We list these properties in Table \ref{diskmod}.

\begin{figure*}[htbp] 
   \centering
   \includegraphics[width=6in]{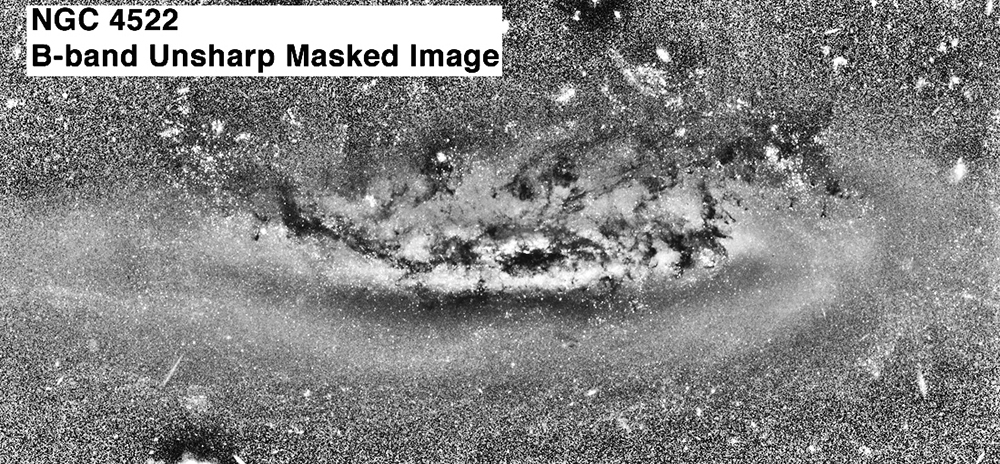} 
   \caption{NGC 4522 B-band unsharp masked image.  Unsharp masking removes large-scale luminosity gradients and highlights smaller dust clouds with interesting morphologies.}
   \label{4522unsh}
\end{figure*}

\begin{figure*}[htbp] 
   \centering
   \includegraphics[width=6in]{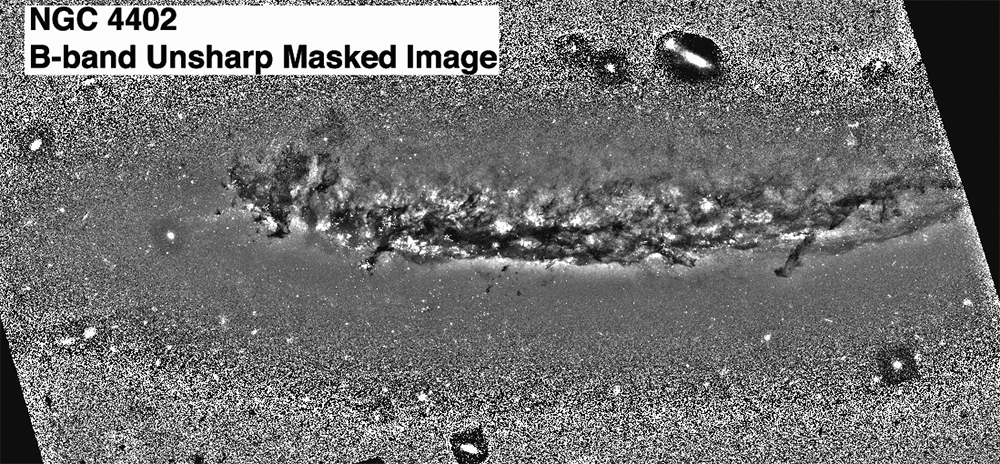} 
   \caption{NGC 4402 B-band unsharp masked image.}
   \label{4402unsh}
\end{figure*}

The unsharp masking procedure consists of convolving an image by a Gaussian kernel, then subtracting the convolved image from the original image at the same wavelength.  This has the effect of removing luminosity variations with spatial scales comparable to the size of the Gaussian kernel, which highlights smaller-scale variations in luminosity such as the dust clouds we examine in this paper.  For both galaxies, we used a Gaussian kernel with $\sigma$ = 4$\arcsec$ to produce unsharp masked images (Figures \ref{4522unsh}, \ref{4402unsh}).  The technique results in oversubtraction near compact, bright sources such as background galaxies (particularly in Figure \ref{4402unsh}), but the dust features in the galaxies are not located near such sources and do not suffer from this effect.

The unsharp masked images are useful for two main reasons.  The first is that all of the isolated dust clouds are visible at the same time, because the radial variation in the stellar disk's surface brightness has been removed.  The second is    
that unsharp masking reveals substructure in the extraplanar dust near the center of the galaxy, which would otherwise be difficult to discern due to the brightness of the nucleus.  The large complexes of extraplanar dust near the galaxies' nuclei will be further examined in Paper II.  For quantitative measurements of dust properties in the rest of this paper, we use the HST V-band images except where otherwise noted.

\begin{table}
\caption{NGC 4522 Model Parameters - V-Band}
\label{diskmod}
\begin{tabular}{|l|l|}
\hline
\hline
Position Angle & 31.5 $\degree \pm$ 2 $\degree$\\
Inclination & 78.0 $\degree \pm$ 0.5 $\degree$\\
Center RA (2000) & $12^{\operatorname{h}}33^{\operatorname{m}}39^{\operatorname{s}}.7$ $\pm$ 0.99 $\arcsec$\\
Center Dec (2000) & +09$\degree$10$\arcmin$27$\arcsec$ $\pm$0.50 $\arcsec$\\
\hline
\end{tabular}\\
\end{table}

\section{Results}
\label{results}

\begin{figure*}[htbp] 
   \centering
   \includegraphics[width=5.5in]{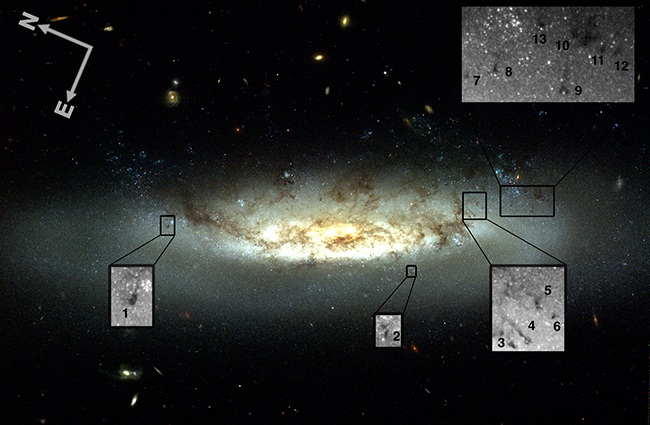} 
   \caption{NGC 4522 HST BVR image with insets of isolated dust clouds in V-band.}
   \label{4522atlas}
\end{figure*}

\begin{figure*}[htbp] 
   \centering
   \includegraphics[width=5.5in]{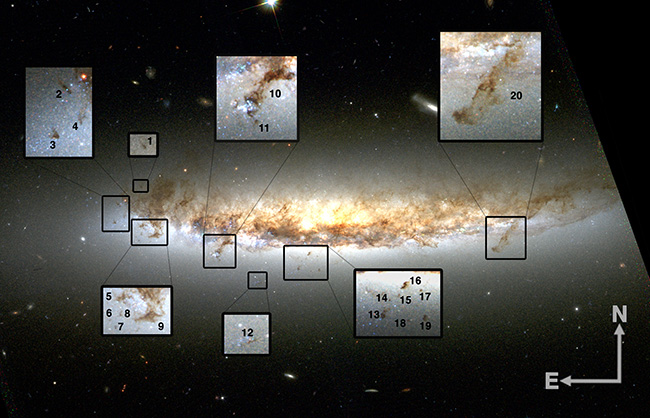} 
   \caption{NGC 4402 HST BVR image with insets of isolated dust clouds.}
   \label{4402atlas}
\end{figure*}

\begin{figure}[htbp] 
   \centering
   \includegraphics[width=3in]{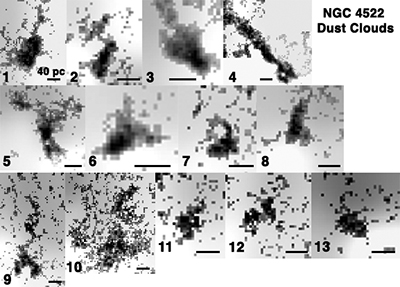} 
   \caption{NGC 4522 dust clouds from V-band images, but showing only pixels more than 1$\sigma$ fainter than the local background, overlaid on a smoothed fit to the local background.}
   \label{4522clouds}
\end{figure} 

\begin{figure}[htbp] 
   \centering
   \includegraphics[width=3in]{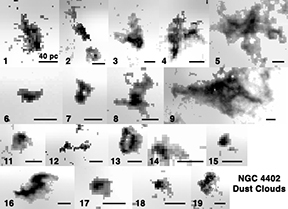} 
   \caption{NGC 4402 smaller dust clouds from V-band images, showing only pixels more than 2$\sigma$ fainter than the local background, overlaid on a smoothed fit to the local background.}
   \label{4402clouds}
\end{figure}

\begin{figure}[htbp]
 \centering
   \includegraphics[width=3in]{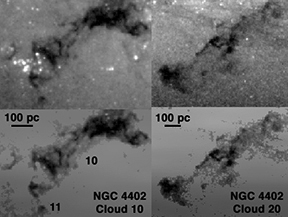} 
   \caption{NGC 4402 larger dust filaments in V-band.  The top panels are the original images, and the bottom panels show only pixels more than 2$\sigma$ fainter than the local background, overlaid on a smoothed fit to the local background.}
   \label{4402cloudsbig}
\end{figure}

\subsection{Dust Cloud Identification and Measurement}
\label{cloudident}
Both galaxies have many small, discrete dust clouds beyond their main dust lanes, and we seek to identify the clouds and measure their properties in a self-consistent way.  Color images of NGC 4522 and NGC 4402 are shown in Figures \ref{4522atlas} and \ref{4402atlas} respectively, with insets showing the dust clouds.  Although the unsharp masked images aided in the identification of some of the isolated dust clouds, our detection criteria are based on the V-band images, and those are the images we measured to determine cloud sizes and masses unless otherwise noted.  
      We define a ``cloud" as an area of the image beyond the main dust truncation boundary (where the dust goes from a continuous distribution to a few discrete clouds, discussed further in Section \ref{tzdef}), where at least one pixel has a surface brightness $> 4\sigma$ below the local background, and where at least 4 contiguous pixels are $> 3\sigma$ below the local background.  Our detection limit is roughly a cloud diameter of 2-3 pixels, or 8-12 pc, and most of the detected clouds range in width from 20 - 80 pc.  Each cloud is located in a ``region" that is comprised of the cloud and the local background.  
      
      For each region, we measure the local background in several spots near the dust cloud that appear to be free from dust extinction or individual bright stars.  Because there is a large gradient in the background surface brightness for some of the clouds, we interpolate linearly between the measured background points to generate an estimate of the background surface brightness value for each pixel in the cloud (i.e., the surface brightness we would observe from the galaxy if the obscuring cloud were not along our line of sight).  Each region with its simulated background and detected dust pixels are shown in Figures \ref{4522clouds}, \ref{4402clouds}, and \ref{4402cloudsbig}.  
      
                 In order to determine which pixels should be considered part of each cloud, we calculate $\sigma$, the standard deviation of the local background surface brightness.  The value of $\sigma$ is higher in NGC 4522 than in NGC 4402 due to the fact that the outer stellar disk in NGC 4522 is younger and the stellar light distribution is less smooth.  In order to calculate cloud mass and radius consistently, we measure the clouds down to similar minimum V-band dust extinction values  (Tables \ref{mass4522} and \ref{mass4402}).  To achieve this, we consider all of the contiguous pixels with surface brightness $>1\sigma$ below the local background to be part of the cloud for NGC 4522, and $>2\sigma$ below the local background for NGC 4402.  
      
      We can calculate a lower limit on the mass of each cloud using a technique similar to \citet{howk97} and \citet{thompson04}.  They use the relationship between dust extinction and gas column density in the Milky Way \citep{bohlin78} to estimate the masses of dust clouds in nearby galaxies.  The apparent dust extinction at a given wavelength $\lambda$ is expressed as the quantity $a_{\lambda}$.  The apparent extinction can be calculated using 
      \begin{equation}
      \label{littlea}
      a_{\lambda} = -2.5\operatorname{log}(S_{dc,\lambda} / S_{bg,\lambda})
      \end{equation}
       where $S_{dc,\lambda}$ is the surface brightness towards the dust cloud, and $S_{bg,\lambda}$ is the local background surface brightness.  The apparent extinction $a_{\lambda}$ does not necessarily match the true extinction, $A_{\lambda}$ because some of the light we interpret as ``background" may actually be located in front of the cloud.  The measured cloud-to-background surface brightness ratio can be expressed using the quantity x, the fraction of light emitted in front of the cloud, as 
       \begin{equation}
       \label{xeq}
       S_{dc,\lambda} / S_{bg,\lambda} = x + (1-x)e^{-\tau_{\lambda}}
       \end{equation}
        where $\tau_{\lambda}$ is the extinction optical depth through the cloud at a given wavelength.  
        
        Combining Equations \ref{littlea} and \ref{xeq} with x=0, we see that the true extinction $A_{\lambda}=1.086\tau_{\lambda}$.  We can then find the relationship between the observed and true extinctions, which is 
        \begin{equation}
        \label{botha}
         a_{\lambda} = -2.5 \operatorname{log}(x + (1-x)e^{-A_{\lambda}/1.086})         
        \end{equation}
       We see that foreground light can have a significant effect on the observed extinction - if 50\% of the assumed background light is actually in the foreground, the observed extinction $a_{\lambda}$ will be about 50\% lower than the actual extinction $A_{\lambda}$.

In this work, we assume that all galactic light is behind the dust features, meaning $a_{\lambda} = A_{\lambda}$, recognizing that the resulting column density estimates will be lower limits.  We can then use the correlation between dust extinction and gas column density in the Milky Way \citep{bohlin78}, 
\begin{equation}
\label{nh}
N_H = 5.8\times10^{21}\,E(B-V)
\end{equation}
where $N_H = N(\mathrm{HI}) + 2 N(\mathrm{H}_2)$, to estimate a lower limit for the total hydrogen column density of the gas in a given cloud.  We can use a Cardelli et al. (1989) extinction law, which describes interstellar extinction with the parameter $R_V$ $\equiv \frac{A_V}{E(B-V)}$.  Observed lines of sight in the Milky Way are well-described by this single-parameter relationship with a mean value of $R_V = 3.1$, with variations related to the size of the dust grains in the cloud.  Dust grains scatter wavelengths of light smaller than their diameters, so if a dust grain is large enough, it will scatter red and blue light equally well, resulting in ``grey" extinction rather than extinction with reddening.  Large dust grains are found preferentially in dense regions \citep[e.g.,][]{cardelli89}.  

We assume a Milky Way average $R_V$ value rather than using the $R_V$ we measure for each cloud in order to be consistent with previous works using this technique  \citep[e.g.,][]{howk97,thompson04}, and because the $R_V$ values we measure in the galaxies are roughly consistent with this value.  We measure average $R_V$ values in NGC 4522 and NGC 4402 of 3.3 and 2.3, respectively, and if we had used those values, our $N_H$ values would have differed by up to 30\%, which is within the expected errors of this work.  Additionally, our measured $R_V$ values may differ somewhat from the actual values due to the effects of foreground light (discussed above).       

Assuming a Milky Way $R_V$ gives the relationship $N_H = 1.9 \times 10^{21} A_V[\mathrm{cm}^{-2}]$, which we use in our calculations.  Varying $R_V$ among the range of observed values in the Milky Way ($R_V$ = 1 - 5, which is similar to the range of variation within the two galaxies) can change the $N_H$ estimate by a factor of $\sim$2.  The relationship between $A_V$ in magnitudes and the ratio of cloud to background flux is approximately linear for the flux ratios in the observed clouds, for $A_V$ values from $\sim$0-0.7, meaning that the relationship between $A_V$ and $N_H$ is approximately linear as well.
    
This approach provides a lower limit on the cloud mass for three reasons.  The first reason is that it is unlikely that all of the galaxy's light in the direction of the cloud is actually behind it - since there is probably some light along the line of sight in front of the cloud, the measured extinction will be lower than the true extinction.  The second is that extinction at optical wavelengths saturates at around $A_V \sim$1, so the dust and gas density can be higher than what we infer for the darkest pixels in some of the clouds, causing us to underestimate the mass in these areas.  The third reason, related to the second, is that substructure on scales too small to resolve may lead to an underestimate of the true dust content within each pixel, because we are able to measure only an average extinction for each pixel. 

\begin{figure}[htbp] 
   \centering
   \includegraphics[width=3.5in]{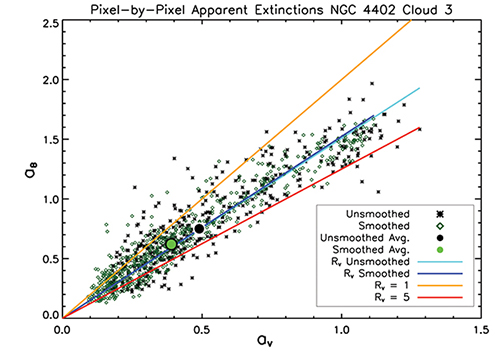} 
   \caption{Pixel-by-pixel apparent extinctions in NGC 4402 Cloud 3.  Smoothing the image (green diamonds) did not yield a significantly different relationship from the original image (black asterisks).  The large filled circles show the cloud averages for the smoothed and original images.  The smoothed value of $R_V$ is 1.91 and the unsmoothed value is 1.97.  Lines indicating $R_V$=1 and $R_V$=5 mark the approximate extreme $R_V$ values in lines of sight observed in the Milky Way.  This figure is discussed in Section \ref{cloudident}.}
   \label{avabplotcloud3}
\end{figure}

We calculate $a_V$ for each individual pixel, then sum the number of hydrogen nuclei estimated for each pixel whose brightness is less than a given limit below the local background.  The total cloud mass is the sum over all the pixels in the cloud.  We also tried this technique with images of the clouds that had been smoothed using a 3x3 box to produce a moving average of the pixel values, in order to see how our technique depends on resolution and to increase the signal-to-noise for some clouds.  We include an example of the pixel-by-pixel $a_V$ measurements for one of the clouds, NGC 4402 Cloud 3 (Figure \ref{avabplotcloud3}).  We measured both a smoothed (green) and unsmoothed (black) image of the cloud.  While the pixel-to-pixel scatter is significant, the nearly linear relationship between $a_V$ and $a_B$ values is clear, and the difference in $R_V$ value is negligible for the smoothed and unsmoothed images of most clouds.  

Smoothing the images of the clouds was sometimes necessary when the unsmoothed images did not have sufficient signal-to-noise to discern a tight relationship between $a_V$ and $a_B$.  This tended to happen in clouds with low-surface-brightness backgrounds.  For this reason, we smoothed NGC 4402 Cloud 1 and NGC 4522 Clouds 10-13 with the same method shown in Figure \ref{avabplotcloud3}.  Smoothing systematically changes the quantities we measure.  In clouds with good signal-to-noise, smoothing lowered the average $a_V$ value by $\sim20\%$, so we correct the smoothed value by a corresponding amount for the five smoothed clouds, giving the final smoothed, corrected values in Figure \ref{avabplot} and Tables \ref{mass4522} and \ref{mass4402}.  Smoothing does not, on average, significantly change the measured cloud mass, so the cloud masses in the tables are from the unsmoothed images.  Smoothing increases the clouds diameters by roughly 5-10\% in the NGC 4522 clouds and  20\% in NGC Cloud 19.  We give the unsmoothed cloud size measurements in the tables because the most of the clouds appear to have well-defined edges, so we believe the unsmoothed size measurements are more accurate.   

The derived cloud masses are listed in Tables \ref{mass4522}  and \ref{mass4402}.  The length is the longest dimension of the cloud, and the width is the cloud's greatest extent along a line perpendicular to the longest dimension.  The cloud's axial ratio is the length divided by the width.  For the more elongated clouds, we measure the position angle of the longer axis, counterclockwise from north (Tables \ref{mass4522} and \ref{mass4402}).

\subsubsection{Analysis of Cloud Properties}
\label{cloudmasses}

To verify our extinction measurements, we confirm that our clouds' wavelength-dependent extinction properties are consistent with known relationships.  We compare the apparent extinction in the B and V bands to see if they have a relationship consistent with a \citet{cardelli89} extinction law (see Section \ref{cloudident}).

\begin{figure}[htbp] 
   \centering
   \includegraphics[width=3in]{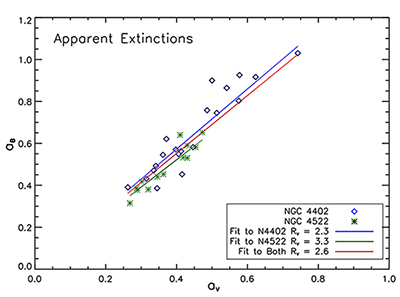} 
   \caption{NGC 4402 and NGC 4522 dust clouds - plot of apparent extinctions in the B and V bands (detection levels of 1$\sigma$ for NGC 4522 and 2$\sigma$ for NGC 4402).  The blue, green, and red lines show fits to NGC 4402, NGC 4522, and both galaxies, respectively.  This plot shows that the $a_V$ and $a_B$ values we measure are consistent with a \citet{cardelli89} extinction law.  This figure is discussed in Section \ref{cloudmasses}.}
   \label{avabplot}
\end{figure}

The majority of the clouds in our galaxies fall within the continuum of wavelength-dependent dust extinction laws that are observed along various lines of sight in the Milky Way (e.g., \citealt{valencic04}, which surveys a range of environments within the Milky Way), as well as galaxy-wide averages in two nearby edge-on spirals whose extraplanar dust has been studied using methods similar to the ones in this paper, NGC 891 ($R_V$=3.6; \citealt{howk97}) and NGC 4217 ($R_V$=3.1; \citealt{thompson04}).  As shown in Figure \ref{avabplot}, we find a linear relationship between $a_v$ and $a_b$ for the two galaxies, with somewhat different $R_V$ values for the two.  The galaxy-wide $R_V$ in NGC 4402 is 2.3, and the value in NGC 4522 is 3.3.  These are both within the range of typical values along lines of sight in the Milky Way.  The dust clouds in NGC 4522 have extinction values at the lower end of the range and lack the higher-extinction clouds present in NGC 4402.  Most of the higher-extinction clouds in NGC 4402 are spatially larger than the typical cloud sizes in NGC 4522 and are very highly-obscuring.  

The difference in $R_V$ values between the two galaxies is only marginally significant, as can be seen in Figure \ref{avabplot}, and both are close to the standard Milky Way value of $R_V$= 3.1 that we use for our calculations.  The systematically higher $R_V$ values in NGC 4522 may mean that the dust grains in the clouds are larger, on average, than the grains in the clouds in NGC 4402.  This is plausible if the clouds we see in NGC 4522 have been stripped down to their densest cores fairly recently, while the clouds in NGC 4402 still contain more diffuse material.  There is some evidence that NGC 4522 has been stripped more rapidly and recently than NGC 4402 (Section \ref{timescales}).  However, a systematic increase in $R_V$ values can also occur if a significant fraction ($\sim$0.4) of starlight is emitted in front of the cloud along our line of sight (see Howk \& Savage 1997, Figure 6).  Dust grain size and foreground light probably both contribute to the scatter in Figure \ref{avabplot}.  

\begin{figure*}[htbp] 
   \includegraphics[width=7in]{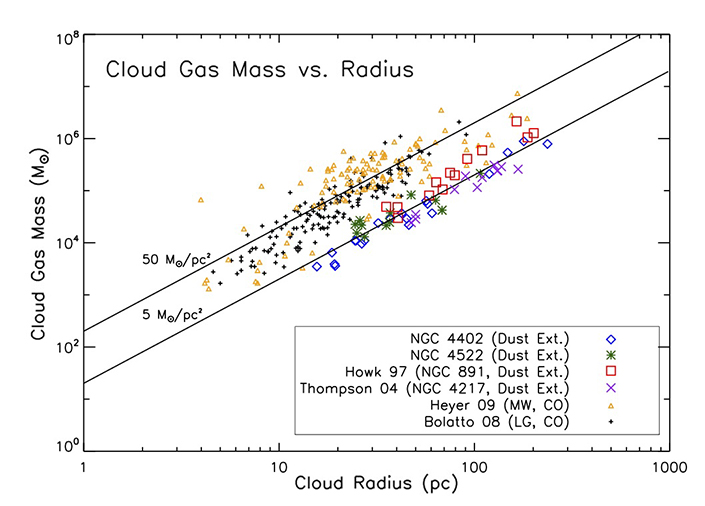} 
   \caption{Cloud gas masses and radii in NGC 4402, NGC 4522, and comparison galaxies.  Smaller symbols are for CO measurements in the Milky Way \citep{heyer09} and local group spirals and dwarfs \citep{bolatto08}.  Larger symbols are from dust extinction-based measurements in NGC 4522 and NGC 4402 (this work), NGC 891 \citep{howk97} and NGC 4217 \citep{thompson04}.  The lines are for surface densities of 5 and 50 M$_{\odot}$ pc$^{-2}$.  This figure is introduced in Section \ref{cloudmasses}.}
\label{gmcmasses}
\end{figure*}

In Figure \ref{gmcmasses}, we compare our dust extinction-based gas masses and radii with those measured by \citet{howk97} and \citet{thompson04} in nearby spirals NGC 891 and NGC 4217.  We also plot gas masses and radii based on CO observations for clouds in the Milky Way and Local Group galaxies (Heyer et al. 2009, Bolatto et al. 2008).  After comparing observed clouds with the isodensity lines at 5 and 50 M$_{\odot}$ pc$^{-2}$, it is clear that the average density of clouds observed in CO varies a bit with mass, tending to decrease at the low end.  For our purposes, this is a second-order effect, since the clouds we observe have higher masses with less variation in the corresponding CO density.  To first order, the clouds observed via optical dust extinction have a roughly constant surface density, which differs by a factor of $\sim$10 from that of the clouds observed in CO.  We believe that similar clouds are being studied - the factor of 10 difference in mass for clouds with a given radius is likely due to the fact that the dust extinction technique underestimates the cloud mass for the three reasons discussed in the previous section.  The difference between dust extinction and CO masses is highlighted in the 
``measured" and ``probable" mass columns in Tables \ref{mass4522} and \ref{mass4402}.

\subsubsection{Cloud Locations Within the Galaxy}
\label{cloudloc}

\begin{figure*}[htbp] 
   \centering
   \includegraphics[width=6.5in]{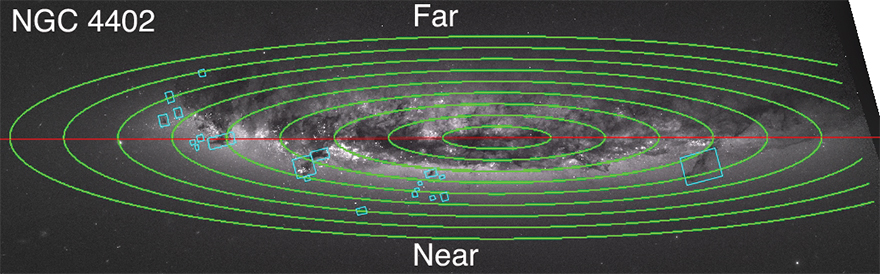} 
   \caption{NGC 4402 dust cloud locations (blue boxes, whose sizes correspond with the cloud sizes), with ellipses which indicate the disk-plane distance (1, 2, 3...kpc) to the galaxy center (same as the distance labeled as $d_{disk}$ in Figure \ref{cloudnear}).  The red line indicates the galaxy's major axis, which demarcates the near and far sides of the galaxy.}
   \label{4402radii}
\end{figure*}

\begin{figure*}[htbp] 
   \centering
   \includegraphics[width=6.5in]{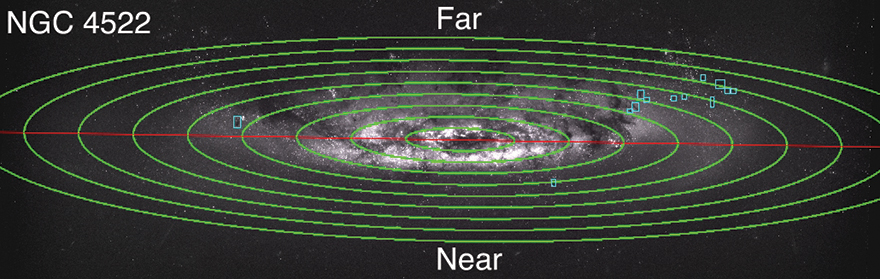} 
   \caption{NGC 4522 dust cloud locations (blue boxes, whose sizes correspond with the cloud sizes), with ellipses which indicate the disk-plane distance (1, 2, 3...kpc) to the galaxy center (same as the distance labeled as $d_{disk}$ in Figure \ref{cloudnear}).  The red line indicates the galaxy's major axis, which demarcates the near and far sides of the galaxy.}
   \label{4522radii}
\end{figure*}

\begin{figure}[htbp] 
   \centering
   \includegraphics[width=3in]{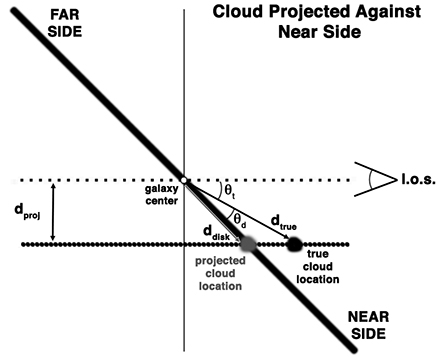} 
   \caption{Geometry of a dust cloud projected against the near side of the galaxy showing that $d_{true} \geq d_{disk}$ for strongly extincting dust clouds.  $d_{disk}$ is the galactocentric radius if a cloud is in the disk plane, and $d_{true}$ is the true distance to the center of the galaxy regardless of whether the cloud is in the disk plane.}
   \label{cloudnear}
\end{figure}

\begin{figure}[htbp] 
   \centering
   \includegraphics[width=3in]{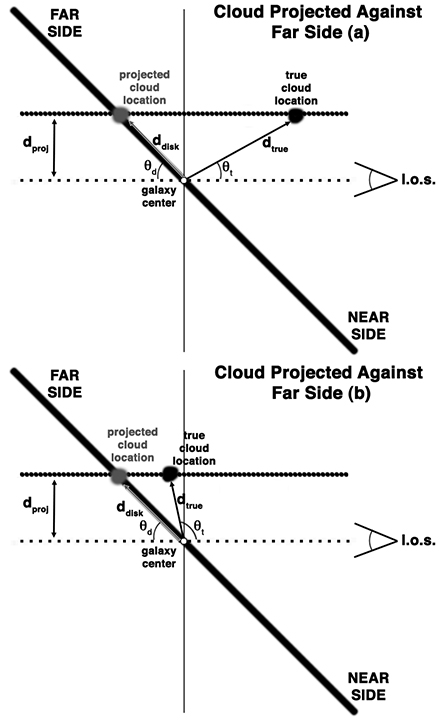} 
   \caption{Geometry of a dust cloud projected against the far side of the galaxy showing two scenarios in which the cloud is at the same projected location in the disk plane but at different distances from the galaxy center.  In a), $d_{true} > d_{disk}$, and in b), $d_{true} < d_{disk}$.}
   \label{cloudfar}
\end{figure}

It is important to determine whether the decoupled clouds are still in or near the disk plane, or whether they are above the disk plane.  We can constrain the clouds' three-dimensional locations based on their projected locations, the galaxy's inclination, and their visibility as extincting clouds.  The clouds' projected locations within the galaxies are shown in Figures \ref{4402radii} and \ref{4522radii}.  The green ellipses in these figures indicate galactocentric radii in the disk plane and are spaced at 1 kpc intervals from the galaxy center.  The ellipses' locations are calculated based on the observed inclination and position angle of each galaxy.  As an initial estimate of the clouds' locations, we can calculate each cloud's distance from the galaxy center (labeled as $d_{disk}$ in Figures \ref{cloudnear} and \ref{cloudfar}) if the cloud were located in the disk plane, and we list these in Tables \ref{mass4522} and \ref{mass4402}.  As we discuss below, their actual locations may differ from this estimate.

In this discussion, we distinguish between the near versus far side of the galaxy (closer to us or further from us than the major axis and galaxy center) and above versus below the disk (closer to us or further from us than the disk midplane).  The constraints on cloud location are based on the assumption that we can only detect dust clouds in certain parts of the galaxy: a dust cloud must be located between us and a significant fraction of the disk light in order to cause noticeable extinction.  Therefore, the clouds we detect must either reside in the disk plane on the near side or above (downstream of) the disk (see also Figure \ref{geom} and the accompanying text in Section \ref{remmass}).  Any transition zone disk-plane clouds on the far side of the galaxy would not be highly visible in dust extinction due to two effects.  The first is the well-known near side-far side dust obscuration difference \citep{slipher17,hubble43,devauc58} - dust clouds near the disk midplane on the far side of a galaxy are less obscuring because there are more stars in front of the clouds.  The second is that transition zone clouds on the far side of the galaxy may also be obscured by stripped extraplanar dust in front of them.  Because of these effects, we believe the clouds we detect are either on the near side in or above the disk plane, or on the far side above the disk.        

We illustrate possible cloud viewing geometries in Figures \ref{cloudnear} and \ref{cloudfar} depending on whether the cloud is projected against the near or far side of the galaxy.  Clouds projected against the near side may be located in or above the disk midplane, so when a cloud is projected against the near side of the galaxy (Figure \ref{cloudnear}), the cloud's true distance from the galaxy center ($d_{true}$ in the figure) must be greater than or equal to the projected distance from the center ($d_{proj}$).  The true distance is also greater than or equal to the distance between the galaxy center and the cloud's apparent location projected onto the disk plane corrected for the inclination of the galaxy (cloud location $d_{disk}$, also indicated by the 1-kpc ellipses in Figures \ref{4402radii} and \ref{4522radii}).  

If a cloud is projected against the far side of the galaxy, where we are unable to detect dust clouds in the disk plane, the cloud's actual location must be above the disk plane.  The clouds projected against the far side could be anywhere along the line of sight - either on the near side (closer than the galaxy center) and far above the disk plane, as shown in the top panel of Figure \ref{cloudfar}, or on the far side (further than the galaxy center) above the disk plane, as shown in the bottom panel of Figure \ref{cloudfar}.  Thus, when a dust cloud is projected against the far side, $d_{true}$ may be either larger (Figure \ref{cloudfar}a) or smaller (Figure \ref{cloudfar}b) than $d_{disk}$.  In this work, we focus on the survival of dense clouds that have remained in the disk.  Therefore, we analyze the locations of only the near-side clouds, since these clouds could be in or near the disk plane - their $d_{disk}$ is a lower limit on $d_{true}$.  We include $d_{disk}$ for clouds projected against the far side in Tables \ref{mass4522} and \ref{mass4402} with asterisks, and we further discuss them in Section \ref{farside}.


\begin{table*}
\caption{NGC 4522 Clouds - Observed and Derived Properties \label{mass4522}}
\begin{tabular}{|c|p{1.5cm}|p{1.7cm}|p{1.7cm}|p{1.5cm}|p{1.5cm} |p{1.5cm}|p{1.5cm}|p{2cm}|}
\hline
Feature & $d_{disk}$ (kpc) & Measured M$_H$(M$_\odot$)& Probable M$_H$(M$_\odot$)& length (pc) & width (pc)& avg $a_v$ & min $a_v$ & PA\\
\hline
       1  &     4.3      &    8.3$\times10^{4}$ &  8.3$\times10^{5}$  & 150  &  60 &      0.44  &  0.16  &   283$\degree \pm 6\degree$ \\
       2  &      4.0     &    2.2$\times10^{4}$ &  2.2$\times10^{5}$  &   100  &  50 &        0.29  &  0.13  &  279$\degree \pm 7\degree$     \\
       3  &      4.2*     &       3.7$\times10^{4}$ &  3.7$\times10^{5}$  &   110   &  50 &     0.42  &  0.14  &  347$\degree \pm 13\degree$    \\
       4  &      4.5*     &      7.9$\times10^{4}$ & 7.9$\times10^{5}$   &   310   & 60  &     0.28  &  0.09  &  346$\degree \pm 4\degree$   \\
       5  &     5.4*      &      4.2$\times10^{4}$ &  4.2$\times10^{5}$    &   170  & 110  &     0.28  &  0.08  &  -   \\
       6  &     5.2*      &      1.3$\times10^{4}$ &  1.3$\times10^{5}$   &   60   & 50  &      0.34  &  0.10  &   -  \\
       7  &      5.6*     &      1.5$\times10^{4}$ &  1.5$\times10^{5}$  &   50   &  50 &     0.26   &  0.11  &   -  \\
       8  &     5.8*      &     2.6$\times10^{4}$ & 2.6$\times10^{5}$   &   90    &  30 &      0.39  &  0.13  & 296$\degree \pm 11\degree$   \\
       9  &     6.1*      &     6.5$\times10^{4}$ & 6.5$\times10^{5}$   &   200    &  80 &       0.34   &  0.14  & 294$\degree \pm 7\degree$    \\
      10  &     7.4*      &    2.1$\times10^{5}$ &  2.1$\times10^{6}$  &   230    &  200 &     0.43**  &  0.21  &   -   \\
      11  &     7.0*      &    2.2$\times10^{4}$ &  2.2$\times10^{5}$  &   60   &  40 &    0.43**   &  0.20  &  -   \\
      12  &    7.0*       &      2.5$\times10^{4}$ &  2.5$\times10^{5}$  &   90   &  60 &     0.32**   &  0.19  &  -   \\
      13  &     7.0*      &    2.2$\times10^{4}$ & 2.2$\times10^{5}$   &     70   &  40 &     0.41**  &  0.22  &  -   \\
\hline
\end{tabular}
\\
\\
Measurements are of pixels with a significance level of at least 1$\sigma$ above background, and include position angles for elongated clouds.  The ``probable" mass column includes an additional factor of 10 based on CO measurements of similarly-sized clouds.  The projected galactocentric radii $d_{disk}$ of clouds projected against the far side of the galaxy are marked with a *.  The $a_V$ values derived from smoothed images (then multiplied by 1.25 to correct for reductions due to the smoothing process) are marked with a **.
\end{table*}

\begin{table*}
\caption{NGC 4402 Clouds - Observed and Derived Properties \label{mass4402}}
\begin{tabular}{|c|p{1.5cm}|p{1.7cm}|p{1.7cm}|p{1.5cm}|p{1.5cm} |p{1.5cm}|p{1.5cm}|p{2cm}|}
\hline
Feature & $d_{disk}$ (kpc) & Measured M$_H$(M$_\odot$)& Probable M$_H$(M$_\odot$) & length (pc) & width (pc) & avg $a_v$ & min $a_v$ & PA\\
\hline
1        &     8.3*      &   2.4$\times10^{4}$ & 2.4$\times10^{5}$   &  90     & 50  &       0.50**    & 0.29    &  37$\degree \pm 5\degree$  \\
  2      &    7.2*       &   3.7$\times10^{4}$   &  3.7$\times10^{5}$   &  170   &  40 &     0.54   &  0.17  &  40$\degree \pm 4\degree$  \\
  3       &    6.4*       &   5.6$\times10^{4}$   &  5.6$\times10^{5}$    &  130   &  100 &      0.45  &  0.15  &  57$\degree \pm 20\degree$ \\
 4       &   6.4*        &   3.7$\times10^{4}$  &  3.7$\times10^{5}$    &  140      &  110 &       0.33    &  0.12  &  -  \\
5       &    5.5        &  2.1$\times10^{5}$  &  2.1$\times10^{6}$     &  260   &  220 &       0.50   &  0.15  &  -   \\
6        &    5.6       &    3.5$\times10^{3}$   & 3.5$\times10^{4}$     &  40     &  20 &       0.31    &  0.17  &  -  \\
 7      &    5.6       &   9.6$\times10^{3}$   &  9.6$\times10^{4}$   &   60     &  50 &       0.43    &    0.15&  -   \\
 8       &   5.5        &   2.2$\times10^{4}$  & 2.2$\times10^{5}$     &  100     & 90  &       0.33   &  0.13  &  -   \\
  9      &     5.1      &   5.4$\times10^{5}$   &  5.4$\times10^{6}$    &  420   & 210  &      0.56  &  0.10  &  -  \\
  10       &   4.5         &  8.9$\times10^{5}$  &  8.9$\times10^{6}$   &  616    & 210   &       0.51   &  0.13  & 331$\degree\pm10\degree$ \\
  11      &    5.3       & 1.1$\times10^{4}$  &   1.1$\times10^{5}$   &  60    &  50 &      0.35  &  0.16   &  -   \\
12        &     7.5      &   1.1$\times10^{4}$  &  1.1$\times10^{5}$    &  90   & 30  &      0.39    &  0.25  &  -   \\
 13        &    5.7       &   2.9$\times10^{4}$   &  2.9$\times10^{5}$    &  80   &  70 &       0.49   &  0.16  &  -  \\
 14       &     5.2      & 1.1$\times10^{4}$  &  1.1$\times10^{5}$   &  50    &  50 &     0.41   &  0.24  &  -  \\
15        &     4.6      &   3.9$\times10^{3}$  &   3.9$\times10^{4}$   &  50   &  30 &    0.33   &  0.13  &  -   \\
 16       &    3.6       &   6.4$\times10^{4}$  &   6.4$\times10^{5}$   &  140    & 90  &       0.52  &  0.12  &  -   \\
17        &    3.9       &   6.5$\times10^{3}$   &  6.5$\times10^{4}$   &  40    & 30  &    0.39   &  0.19  &  -   \\    
 18       &    5.9       &   3.6$\times10^{3}$   &   3.6$\times10^{4}$   &  50   & 30  &      0.25  &  0.14  &  -   \\  
19        &     5.8      &   2.9$\times10^{4}$  &  2.9$\times10^{5}$     &  100     & 80  &       0.40   &  0.14  &  -  \\
 20      &     4.5       &  7.9$\times10^{5}$     &  7.9$\times10^{6}$     &  900   & 250  &  0.35  &   0.12 &  325$\degree \pm 5\degree$  \\ \hline
\end{tabular}
\\
\\
Measurements are of pixels with a significance level of at least 2$\sigma$ above background, and include position angles for elongated clouds.    The ``probable" mass column includes an additional factor of 10 based on CO measurements of similarly-sized clouds.  The projected galactocentric radii $d_{disk}$ of clouds projected against the far side of the galaxy are marked with a *.  The $a_V$ values derived from smoothed images (then multiplied by 1.25 to correct for reductions due to the smoothing process) are marked with a **.

\end{table*} 

\section{Discussion}
\label{discussion}

\subsection{Elongated Dust Cloud Morphology and Orientation}
\label{cloudorient}
Elongated dust clouds can be a direct indicator of active ICM ram pressure.  While not all of the decoupled clouds in the galaxies show the elongation characteristic of active ram pressure, the transition zones (the area beyond the main dust truncation radius, defined in more detail in Section \ref{tzdef}) of both galaxies contain some elongated dust clouds with distinct position angles.  The presence of these elongated clouds shows that their current locations are experiencing active ram pressure as many of the decoupled clouds are being shaped, ablated, and probably destroyed by the ICM wind.

\begin{figure*}[htbp] 
   \centering
   \includegraphics[width=6.5in]{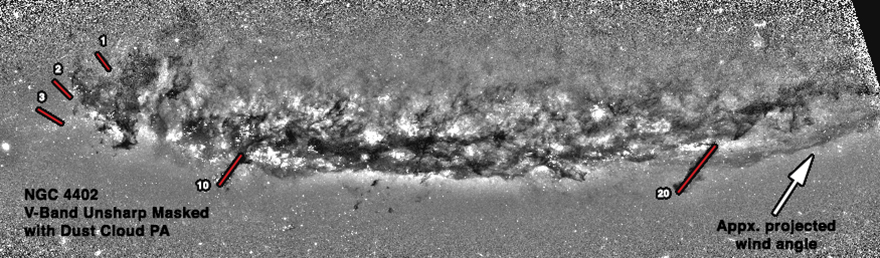} 
   \caption{NGC 4402 V-Band unsharp masked image with elongated dust clouds' approximate position angles.  An arrow indicates the approximate projected ICM wind angle based on the galaxy's HI and radio continuum tails and radio deficit region.}
   \label{4402pa}
\end{figure*}

\begin{figure*}[htbp] 
   \centering
   \includegraphics[width=6.5in]{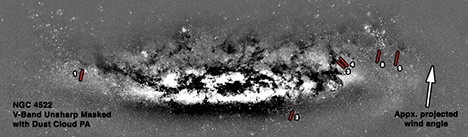} 
   \caption{NGC 4522 V-Band unsharp masked image with elongated dust clouds' approximate position angles.  An arrow indicates the approximate projected ICM wind angle based on the galaxy's HI and radio continuum tails and radio deficit region.}
   \label{4522pa}
\end{figure*}

In Figures \ref{4402pa} and \ref{4522pa}, we show schematics of these clouds' locations within the galaxies.  The red line at each cloud's location indicates the cloud's approximate position angle.  The elongated clouds have a range of morphologies, most with some suggestion of distinct head (denser and/or less elongated) and tail (more diffuse and/or more elongated) components.  
 
We find that in both galaxies, there are clouds beyond the main dust truncation radius that are aligned with each other, and whose position angles are consistent with that of the projected ICM wind, whose direction has been inferred based on the HI and radio continuum tails and the radio deficit regions (see Paper II).  Notably, while there are local systemic variations in the position angle of the dust clouds (e.g., the NE complex in NGC 4402 and NW complex in NGC 4522), the ones oriented in the ICM wind direction are found even on different sides of the galaxies.  A more complete discussion of the patterns of cloud elongation is given in Paper II.  For the present paper, we wish to emphasize only that the cloud elongation indicates that the transition zone clouds are experiencing active ram pressure.

\subsection{Concept of the ``Transition Zone"}
\label{tzdef}
Both galaxies show a distinctive pattern in their outer disks: beyond the well-defined edge of the main dust lane, there is a population of small, discrete dust clouds with virtually no dust extinction between them, and beyond the small clouds the disk is entirely free of visible dust extinction.  We introduce the term ``transition zone" to describe the region of the galaxy within 1 kpc of the disk plane, extending radially from the outer edge of the area in the central disk that is almost completely filled by dust, to the outermost isolated dust cloud.  The location and extent of the transition zone can vary azimuthally within a galaxy.  Throughout large portions of the two galaxies, there is a well-defined boundary inside which the dust distribution is nearly continuous.  Outside of this boundary, however, the disk appears empty of dust extinction except for a number of small, discrete clouds.  Beyond the transition zone, there is no obvious dust extinction, though the stellar disk in both galaxies remains bright enough to detect clouds for distances between 2 and a few kpc.  

We think that the transition zone is a portion of the disk which has been partially, but not completely, stripped of its ISM.    Our interpretation is that the remaining small clouds in the transition zone were once local ISM density peaks, and now all of the less-dense gas around them has been stripped away.  The transition zones of both galaxies appear to be experiencing active ram pressure since they have clouds elongated in the likely ICM wind direction (Section \ref{cloudorient}).  Not all of the discrete small clouds we detect are in the transition zone under this definition.  We do not count clouds projected against the far side of the galaxy towards the total mass in the transition zone since they are not located in the disk plane.  If a cloud could potentially be within $\sim$1 kpc of the disk midplane on the near side of the galaxy, we consider it ``in the disk plane" for the purposes of this study.  We find two clouds in NGC 4522 and 16 clouds in NGC 4402 that meet these criteria.  Note that clouds projected against the near side may actually be above the disk plane - our estimate of the mass in the transition zone includes all of the observed clouds that could potentially be in or near the disk plane.  Our estimate should therefore be considered an upper limit on the mass contained in the remaining decoupled clouds in the disk.  

\begin{figure*}[htbp] 
   \centering
   \includegraphics[width=6.5in]{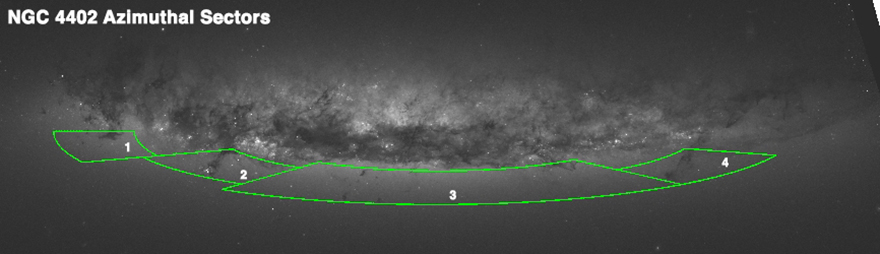} 
   \caption{NGC 4402 transition zone azimuthal sectors, used for calculating remaining mass in transition zone.  This figure is introduced in Section \ref{tzdef}.}
   \label{4402tz}
\end{figure*}

\begin{figure*}[htbp] 
   \centering
   \includegraphics[width=6.5in]{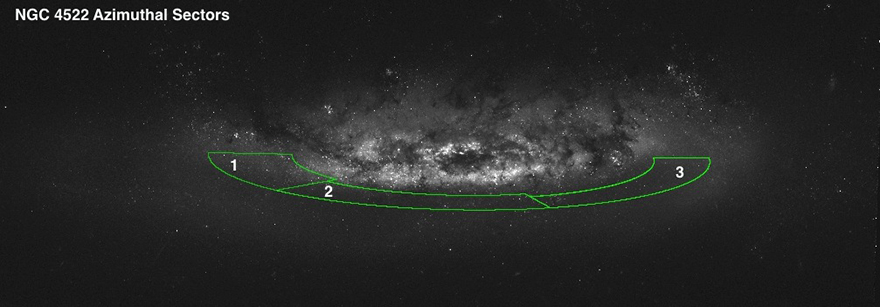} 
   \caption{NGC 4522 transition zone azimuthal sectors, used for calculating remaining mass in transition zone.  This figure is introduced in Section \ref{tzdef}.}
   \label{4522tz}
\end{figure*}

In Figures \ref{4402tz} and \ref{4522tz}, we outline the transition zones of NGC 4402 and NGC 4522.  The radius at which the transition zone begins and ends varies azimuthally throughout the disks of both galaxies, and the radial width of the zone is generally $\sim$1 - 2 kpc.  Because the location and width of the transition zone vary with azimuth, we divide the near side of each galaxy into azimuthal sectors with common inner and outer radii.  In both galaxies, the outer radius is the maximum projected radius of the individual dust clouds within the sector, rounded up to the nearest 0.5 kpc.  The inner radius is the radius at which we detect little or no dust extinction from the galaxy's central, continuous dust distribution in the majority of the sector.  The exact location of the inner radius is clearer in NGC 4402, where the central dust distribution generally has sharp edges.  The edge of the central dust distribution in NGC 4522 is fairly well-defined, although in some parts of the galaxy the dust becomes diffuse at the edges.  Sectors 1 and 3 are free of significant dust extinction apart from one dense cloud each, and in these sectors the inner radius is the radius at which the diffuse dust from the central dust lane seems to end.  Sector 2 does contain a large, mostly diffuse dust complex, but the majority of the sector appears dust-free, and we base our estimate of the inner radius on the parts of the sector outside of the dust complex.    



NGC 4522's Sector 2 is the only portion of the two galaxies that is not well-characterized by our description of the transition zone because the main dust lane does not have a well-defined cutoff.  Rather, the dust extinction gradually becomes more diffuse before tapering off completely.  To approximate the central disk truncation radius, we take the average of the other two sectors that do have clearer boundaries, and for the outer radius, we use the outer radius of the other two sectors.  In this sector, there is a large dust complex with highly-extincting regions connected by more diffuse dust extinction.  In its longest dimension, the complex is about 1.2 kpc long, which makes it about 30$\%$ longer than the large filaments in NGC 4402.  Rather than having the clearly-defined edges of all of the other transition zone dust features in the two galaxies, this complex is very diffuse at its edges and does not have clear boundaries.  We still do not detect any small, discrete clouds in this portion of the transition zone - apart from the large dust complex, this sector does appear to be relatively free of dust extinction beyond the end of the galaxy's main dust lane.  Thus, Sector 2 appears to have been partially stripped, although the remaining dust extinction does not show any of the expected evidence of direct pressure, such as a lack of diffuse material or linear, elongated clouds.  We explore possible explanations for the appearance of Sector 2 in the following section.

In simple, smooth-ISM models of ram pressure stripping of inclined galaxies (e.g. \citealt{roediger06}, \citealt{jachym09}, \citealt{vollmer09}), the azimuthal variation of the dust truncation boundary is strongly affected by the location of the galaxy's leading edge as it moves through the ICM.  In simple models, two important factors determine the asymmetry of the remaining ISM: rotation and shielding \citep{phookun95, roediger06, vollmer01}.  These effects  drive azimuthal asymmetries in the remaining ISM disk, but if the ISM distribution has no substructure,  the truncation radius will still vary smoothly with azimuth.  As time passes, the whole galaxy will rotate through areas of both high and low pressure.  If the timescale for changes in ram pressure strength is long compared to the rotation timescale, this will eliminate any dramatic azimuthal changes in the dust truncation radius. 

Since the transition zone boundaries of NGC 4402 and NGC 4522 do not vary smoothly with azimuth, the effects described above do not provide an adequate explanation for the irregular radial variation of the truncation boundary.  We suggest that pre-existing substructure in the ISM is more important than the distance from the leading edge in determining how deeply a given azimuthal portion of the galaxy is stripped.  More sophisticated simulations of ram pressure stripping \citep[e.g.,][]{tonnesen12} show small, dense clouds left behind after most of the outer disk has been stripped.  The density of the ISM within a galaxy can vary a great deal.  On large scales, there are density contrasts between arm and interarm regions.  ISM density contrasts due to spiral arms and smaller substructure are present in all spiral galaxies whether or not they are undergoing stripping.  We believe ISM substructure is a likely cause of the irregular azimuthal locations of the dust truncation boundaries and the transition zones in NGC 4402 and NGC 4522, as well as the irregular distribution of the surviving clouds.  

\subsection{Remaining Mass in the Transition Zone}
\label{remmass}

\begin{figure}[htbp] 
   \centering
   \includegraphics[width=3.75in]{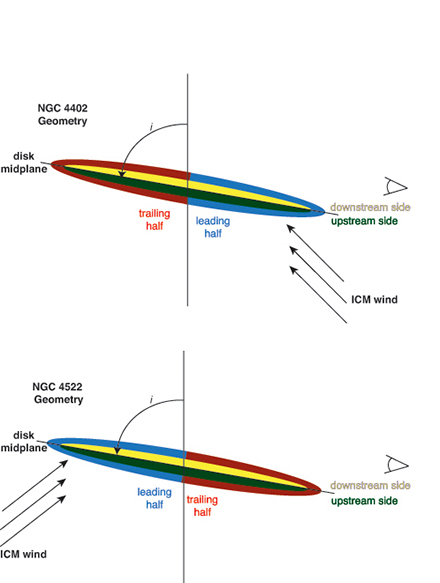} 
   \caption{Viewing geometry of NGC 4402 and NGC 4522, showing an edge-on view of the galaxy where the observer is located to the right of the frame.  The inclination of the stellar disk is $i$. }
   \label{geom}
\end{figure}

In this section, we estimate how much of the original ISM mass remains in the transition zone, the portion of the disk just beyond its main ISM stripping boundary, on the near side of each galaxy.  The remaining mass in the transition zone is contained in the small clouds and large dust plumes described in Section \ref{cloudmasses}, which have decoupled from the lower-density ISM that has already been stripped away.  We confine our discussion to the near side of the galaxy since the clouds projected against the near side may be in or near the disk plane, but the clouds projected against the far side must be extraplanar.  Since we define the transition zone as a part of the disk plane, we include only clouds that may be in or somewhat above the disk plane on the near side of the galaxy.  Some of the clouds projected against the near side may actually be above the disk plane, so our estimate in this section should be interpreted as an upper limit on the mass of decoupled clouds in the disk plane.  Throughout this section, we assume that we can only detect clouds in or in front of the disk plane, and that clouds behind the disk plane will have too much starlight from the galaxy's disk in front of them to be visible (see Figure \ref{cloudnear} for a diagram of the relationship between the clouds and the disk plane).  Although we are unable to detect clouds behind the disk plane, we would not expect to find a significant number of clouds there, because in both galaxies the downstream direction is above the disk (see Figure \ref{geom} for a diagram of the viewing geometry and galaxy motion).  Since the side below the disk plane is more directly exposed to ram pressure, we would expect that it has been more strongly stripped than the visible side.  Therefore, we do not make corrections for clouds located behind the disk plane.  

We assume that the only ISM remaining in the transition zone is contained in the dense, obscuring clouds, since we observe little dust outside the clouds.  Our surface density detection limits vary somewhat by galaxy and radius (see Section \ref{cloudident} for a full explanation of the detection thresholds), but a typical detection limit would be a cloud $\sim$10 pc in diameter with $\Sigma_{\text{HI+H$_{2}$}}$= 2 M$_\odot$ pc$^{-2}$ at 1$\sigma$ significance as measured, or 20 M$_\odot$ pc$^{-2}$ after accounting for the factor-of-ten difference between dust extinction and CO mass estimates.  The significance of the detection increases as the number of pixels in the cloud increases.

We estimate the pre-stripping gas surface density in each sector of the transition zone (the sectors are described in Section \ref{tzdef} and Figures \ref{4402tz} and \ref{4522tz}).  It is well-established that in nearby spiral galaxies that have not been affected by their environments, there is a relationship between the neutral gas surface density  $\Sigma_{\text{HI+H$_{2}$}}$ and the surface brightness of the optical disk (e.g., \citealt{bigiel12}, \citealt{blitz04}).  \citet{bigiel12} show that the undisturbed galaxies in their sample have approximately exponential $\Sigma_{\text{HI+H$_{2}$}}$ profiles with similar scale lengths when the radius is scaled to $r_{25}$ and the surface density is scaled to its value at the point in the galaxy where $\Sigma_{\text{HI}}$= $\Sigma_{\text{H$_{2}$}}$.  Since the relationship does not hold well for galaxies that have undergone ram pressure stripping, we use a simplified equation and assume that the $\Sigma_{\text{HI+H$_{2}$}}$ in each galaxy used to have an exponential distribution with the same scale-length as the galaxy's optical disk.  The observed gas masses come from VLA HI data of both galaxies from the VIVA (VLA Imaging of Virgo in Atomic gas) survey \citep{chung09}, CO data of NGC 4522 from \citet{vollmer08}, and CO data of NGC 4402 from \citet{kenney89}.  We then calculate what the gas distribution in the transition zone would be if the central exponential distribution continued outward, integrating the exponential distribution over the transition zone radii to estimate the ISM mass in each azimuthal sector before it was stripped.  We elect to use simulated HI and CO profiles for the inner galaxy rather than observed ones because the observed maps' resolution ($\sim$1 kpc) is insufficient to explore the $\lesssim$100 pc spatial scales we are interested in.

This method gives a reasonable estimate of the pre-stripping gas mass, but it does have limitations.  The original surface density may have been higher or lower than a smooth radial exponential fit suggests, since a spiral galaxy has higher-density spiral arms and less-dense interarm regions.  In M51, one of the nearest and best-studied spirals, the HI + H$_2$ gas surface density contrast (with H$_2$ measured from $^{12}$CO 2-1) between the arm and interarm regions at a given radius is 2-3 \citep{hitschfeld09} at 11$\arcsec$ (450 pc) resolution.  The same study shows smaller-scale density enhancements up to a factor of $\sim$10 on spatial scales comparable to its resolution limit, probably due to dense molecular clouds.  M51 has unusually strong spiral structure, and the arm-interarm contrast is likely smaller in most galaxies.  If the strong, continuous dust lane at the inner edge of the transition zone is a spiral arm, and the transition zone is an interarm region, then assuming an arm-interarm mass contrast of 2 - 3, the original mass of the transition zone may be a factor of $\sim \sqrt{2}$-$\sqrt{3}$ lower than is predicted by a smooth exponential distribution.  Conversely, if the transition zone was originally a spiral arm, the original mass may be a factor of  $\sim \sqrt{2}$-$\sqrt{3}$ higher than the exponential model.  

\begin{figure}[htbp] 
   \centering
   \includegraphics[width=3in]{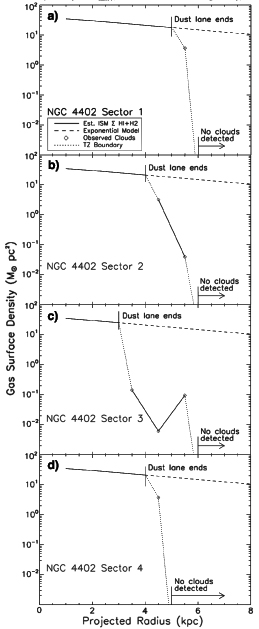} 
   \caption{Gas surface density $\Sigma_{\text{HI+H$_{2}$}}$ of the remaining ISM and modeled pre-stripping mass in NGC 4402 azimuthal sectors.  The observed gas surface density in the transition zone is indicated with diamonds, and the predicted exponential distribution is shown by the dashed lines.  This figure is discussed in Section \ref{remmass}.}
   \label{4402sdmulti}
\end{figure}

\begin{figure}[htbp] 
   \centering
   \includegraphics[width=3in]{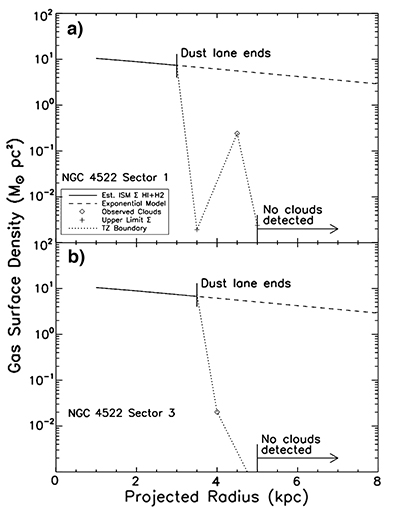} 
   \caption{Gas surface density $\Sigma_{\text{HI+H$_{2}$}}$ of the remaining ISM and modeled pre-stripping mass in NGC 4522 azimuthal sectors.  The observed gas surface density in the transition zone is indicated with diamonds, and the predicted exponential distribution is shown by the dashed lines.  This figure is discussed in Section \ref{remmass}.}
   \label{4522sdmulti}
\end{figure}

The estimated pre- and post-stripping surface densities for NGC 4402 and NGC 4522 are shown in Figures \ref{4402sdmulti} and \ref{4522sdmulti}.  Figure \ref{4402sdmulti} shows all four sectors of NGC 4402 with their modeled pre-stripping and observed post-stripping ISM surface densities and includes the factor of 10 mass correction discussed in Section \ref{cloudmasses}.  The observed surface density in the transition zone, as indicated by a dotted line connecting open diamonds, is the mass of all the small clouds in a 1-kpc radial bin of the sector divided by the surface area of the bin.  The solid and dashed lines represent the surface density predicted by the exponential disk model described above.  The solid line shows points at radii smaller than the transition zone, where most or all of the original ISM is still in the disk.  The dashed line shows what the original exponential distribution would look like if it extended into the transition zone.  The boundaries of the transition zone are marked.  Figure \ref{4522sdmulti} shows the same information for NGC 4522.  Sector 2 of NGC 4522 does not lend itself to this technique since there is a large dust complex rather than discrete clouds with a well-defined truncation boundary.

We then sum the masses of each galaxy's sectors to get the total estimated original mass in its near-side transition zone.  The remaining mass in each sector is the sum of the small clouds and/or large dust plumes contained in the sector.  There is some ambiguity as to whether clouds located just above a galaxy's major axis are located near the disk plane or not, and in these calculations we include them (Cloud 9 in NGC 4402 and Cloud 1 in NGC 4522).

We divide the remaining mass by the estimated original mass to get the fraction of mass remaining in the transition zone at the current stage of stripping.  Both the measured and likely values, the latter incorporating the factor-of-10 correction, are given in Table  \ref{tzprops}.  Even with the larger cloud mass estimate, there is still a small fraction of the original mass remaining - $\sim$6.3\% for NGC 4402, and $\sim$1.3\% for NGC 4522 if we consider only Sectors 1 and 3.  The remaining fraction could be larger or smaller by a factor of $\sim \sqrt{2}$-$\sqrt{3}$ depending on the spiral structure present in the transition zone before stripping, since the transition zone may have been an overdense arm or an under dense interarm region.  Therefore, the remaining fraction could as low as 3.7\% or as high as 11\% for NGC 4402, and the remaining fraction in NGC 4522 could be 0.8\% - 2.4\%.   


As described in the previous section, Sector 2 in NGC 4522 is a special case - it lacks dust extinction over much of its area but still contains one large dust complex.  There are no visible decoupled clouds in Sector 2.  The Sector 2 dust complex has a diffuse, extended morphology unlike that of a decoupled cloud being affected by the ICM wind.  Given how stripped the rest of the transition zone is, it is unlikely to be a portion of the disk plane that has simply been unaffected by ram pressure.  We think the most likely scenario is that the Sector 2 dust complex originated in the disk plane at a larger radius than its current projected one and has been pushed above the disk plane by the ICM wind.  The galaxy contains many other extraplanar clouds downwind of the disk with similar diffuse, extended morphologies that must have originated from material that was stripped from the disk plane, most of which are viewed against the far side of the galaxy.  Although we do not think the Sector 2 cloud meets our criteria for transition zone clouds, there is no way to be sure that it is not in the disk plane.  If it is, then 10\% -- 33\% of the pre-stripping ISM remains in the sector, depending on whether it was an arm or interarm region.  The dust complex covers a similar fraction of the sector's surface area, suggesting that the surface density of this cloud is similar to the pre-stripping azimuthally averaged surface density of this part of the disk.

Our measurements indicate that ram pressure stripping has removed nearly all of the ISM mass in the transition zones of NGC 4522 and NGC 4402.  We note that there is an almost complete lack of diffuse dust clouds in between the more obscuring, discrete clouds in the transition zones of NGC 4522 and NGC 4402.  This seems to show that the primary effect of ram pressure stripping on these galaxies' transition zones has been to remove almost all low density gas outside of the remaining dense cores, which contain a small fraction of the original ISM mass.  We look forward to future high-resolution CO observations of these galaxies to test this prediction.

\begin{table*}[htbp]
   \caption{Derived Transition Zone Properties}
   \begin{tabular}{ lc|c|c|c|c| } 
          & TZ & TZ Total Measured & Pre-Stripping Gas& TZ Total Measured & Likely Mass  \\
   Galaxy& Radius & Cloud Mass & Mass of Sectors & Cloud Mass Remaining & Remaining\\
   		& (kpc) & (M$_{\odot})$ & (M$_{\odot})$ & &\\
   \hline
        NGC 4522 &  3 - 4.5  & 1.1$\times10^5$ & 7.9$\times10^7$ & 0.13\% & 1.3\%\\
        NGC 4402 &  3 - 5.5 & 2.6$\times10^6$ & 4.1$\times10^8$ & .63\% & 6.3\% \\
   \end{tabular}
   \\
   \\
Derived transition zone properties, with ``measured mass remaining" via the dust extinction method, and ``likely mass remaining" based on the factor of 10 underestimate using the dust extinction method (see Section \ref{cloudmasses}).  The estimate of pre-stripping mass is based on an azimuthally-averaged exponential model of $\Sigma_{\text{HI+H$_{2}$}}$, and the total measured cloud mass remaining is based on the TZ total cloud mass and the pre-stripping gas mass of the sectors.
   \label{tzprops}
\end{table*}

\subsection{Star Formation in the Transition Zone}
\label{sftz}
We use ground-based H$\alpha$ + [NII] (hereafter referred to as H$\alpha$) images from the WIYN telescope with 1.0-1.1\arcsec resolution (\citealt{crowl05}, \citealt{kenney99}) to determine whether there is any star formation in the transition zones.  We searched for discrete, compact sources of H$\alpha$ in the transition zones, which we assume are HII regions.  We have the sensitivity, assuming a source size of 1\arcsec \, FWHM, to detect sources with a flux of $8.3\times10^{-17}$ erg s$^{-1}$ cm$^{-2}$ in NGC 4402 and $5.2\times10^{-17}$ erg s$^{-1}$ cm$^{-2}$ in NGC 4522 at 3$\sigma$.  Both galaxies have some diffuse H$\alpha$ emission in their transition zones, but this is likely the product of ionizing photons that have escaped from the star formation regions in the inner galaxy and come in contact with stripped gas above the disk plane.  

\begin{figure}[htbp] 
   \centering
   \includegraphics[width=3.5in]{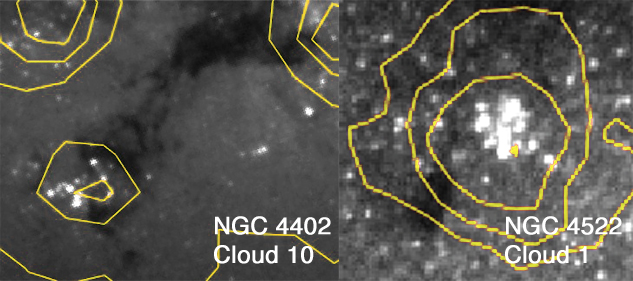} 
   \caption{The two transition zone clouds with H$\alpha$ detections, NGC 4402 Cloud 10 and NGC 4522 Cloud 1, in HST V-band greyscale with H$\alpha$ contours.  Contour levels are 5.4, 7.5, $9.6\times10^{-17}$ erg s$^{-1}$ arcsec$^{-2}$ cm$^{-2}$ in NGC 4402 and 1.6, 3.3, $4.8\times10^{-17}$ erg s$^{-1}$ arcsec$^{-2}$ cm$^{-2}$ in NGC 4522.}
   \label{cloudha}
\end{figure}

We find that each galaxy has one H$\alpha$ region in its transition zone with the compact morphology characteristic of star-forming regions.  In NGC 4522 the region is near Cloud 1, and in NGC 4402 the region is is at the head of Cloud 10, one of the large dust filaments.  Figure \ref{cloudha} shows overlays of H$\alpha$ contours on the HST images for these regions.  There are no transition zone HII regions that are not associated with dust clouds.  The H$\alpha$ region in NGC 4522 accounts for 0.6\% of the galaxy's total H$\alpha$ luminosity, or $8.4\times10^{37}$ erg s$^{-1}$ (Table \ref{hatz}).  The H$\alpha$ region in NGC 4402 contains 0.5\% of the galaxy's H$\alpha$, a luminosity of $4\times10^{38}$ erg s$^{-1}$.

We can estimate the amount of H$\alpha$ that would have been in the transition zone prior to stripping, using a similar technique to our transition zone remaining gas mass estimates.  Galaxies' scale lengths in R-band and H$\alpha$ are generally comparable \citep{koopmann06}, so we can use the R-band scale length and current H$\alpha$ luminosity to derive a simulated exponential distribution extending into the transition zone.  We use the same boundaries for the transition zones as in Section \ref{remmass}.  The observed H$\alpha$ luminosity in the transition zone is 3\% of the likely pre-stripping amount in NGC 4522, and 2\% in NGC 4402.  


Relating an H$\alpha$ luminosity to a star formation rate in the transition zones can only be done approximately due to the small number of HII regions involved.  Using H$\alpha$ to calculate a star formation rate with any accuracy requires a large number of massive stars distributed roughly uniformly in age (e.g., \citealt{kennicutt98}).  The technique depends on statistical averaging over a population with a range of stellar masses and ages, and its accuracy is limited when there are only a few HII regions, as is the case in the transition zones.  The galaxy-wide H$\alpha$-SFR relationship can vary based on the chosen IMF by a factor of $\sim$3, and our estimates are further limited by the small number of HII regions, so we expect our SFR estimates to have errors of a factor of $\sim$several.  If the standard H$\alpha$-star formation rate relationship is applied to the HII regions, then the amount of star formation in the transition zone is 2-3\% of the pre-stripping level.  This is similar to the surviving HI + H$_2$ fraction in the transition zones of 1-6\%.  Given the uncertainty in relating H$\alpha$ luminosity to a star formation rate, we would not expect an agreement between the surviving gas mass and the surviving H$\alpha$ luminosity fraction to greater than a factor of a few, even if the surviving gas mass and SFR fractions are identical.             

It appears that within our detection limits, there is only a limited amount of star formation associated with the decoupled transition zone clouds, since only two of the clouds have associated star-forming regions.  The star formation that we do detect in the transition zones may have been triggered by ram pressure, or it may have occurred in the cloud independent of ram pressure.  If star formation were triggered in the transition zones, we might expect to see a higher surviving fraction of H$\alpha$ than surviving gas mass in these regions.  The percentages, however, are comparable, so ram pressure does not seem to frequently trigger star formation in decoupled clouds like the ones in the transition zones.

\begin{table*}[htb]

\caption{Current and Estimated Pre-Stripping H$\alpha$ Luminosities in NGC 4522 and NGC 4402.}
   \begin{tabular}{ lc|c|c|c|c|c| } 
 
   & Total               & TZ                 & Pre-Stripping &  &  Current  & Pre-Stripping     \\
    Galaxy           &L$_{H\alpha}$ &L$_{H\alpha}$  &TZ L$_{H\alpha}$ & \% Surviving & TZ  SFR   & TZ SFR   \\
                &   (erg s$^{-1}$)   &   (erg s$^{-1}$)  &  (erg s$^{-1}$)   &  &(M$_{\odot}$ year$^{-1}$)   &  (M$_{\odot}$ year$^{-1}$)   \\
   \hline                
        NGC 4522 &  $1.3\times10^{40}$  & $8.4\times10^{37}$ & $2.8\times10^{39}$ &  3\% & $6.6\times10^{-4}$ & 0.02 \\
        NGC 4402 &  $7.9\times10^{40}$ & $4\times10^{38}$ & $1.8\times10^{40}$& 2\%  &0.003 &  0.14\\
    
   \end{tabular}
   
   \label{hatz}
\end{table*}

\subsection{Constraints on Dust Cloud Lifetimes and Timescales for Stripping}
\label{timescales}

If we assume that star formation is quenched when the low-density gas (the gas not contained in decoupled clouds) is stripped, and we can accurately measure the quenching time in a region with decoupled clouds, we can estimate the lifetime of the decoupled clouds.  This is possible in NGC 4522 - the quenching time in Sector 3 of the transition zone is 100$\pm$50 Myr \citep{crowl08}, so decoupled clouds survive at least 100$\pm$50 Myr.  In NGC 4402, \citet{crowl08} measure the quenching time in a region just beyond the transition zone to be $\sim$200 Myr, so the decoupled clouds survive $<$200 Myr.  


The average lifetime of a molecular cloud in a typical star-forming spiral galaxy is estimated to be $\sim$20-40 Myr (e.g. \citealt{miura12}, \citealt{blitz07}, \citealt{kawamura09}) before they are destroyed by star formation activity, although it is likely that some clouds will have longer or shorter lifetimes.  It is possible that the lifetimes of the transition zone clouds are in this range.  The clouds we see now were almost certainly larger at the time of decoupling and have since been ablated by the ICM wind.  We detect only the longest-lived transition zone clouds which had sufficient mass and/or density to outlast any other clouds that may have also decoupled from the ISM.  The quenching time in the transition zone may be more recent at the clouds' locations than where the spectra used to determine quenching times were taken, especially in NGC 4402.  In any case, the margin of error on the quenching time estimates is relatively large.  We do not observe significant star formation associated with most of the transition zone molecular clouds, so it is also possible that they last longer than molecular clouds in typical disks if star formation is suppressed by the other environmental processes taking place.  

The size of the transition zone is related to the stripping rate and cloud lifetimes.  In a simple scenario with a continuous and uniform stripping rate, we can say that the size of the transition zone is equal to the stripping rate times the decoupled cloud survival time.
The only quantity we can directly measure is the size of the transition zone.  If we adopt a cloud survival time, we can estimate the stripping rate, and if we adopt a stripping rate, we can estimate the cloud survival time.

We can adopt cloud survival times based on the quenching times above to estimate stripping rates for the galaxies.  In NGC 4522, the transition zone varies from 0.5 -- 1.5 kpc wide, giving a possible stripping rate range of 3 pc Myr$^{-1}$ -- 30 pc Myr$^{-1}$.  In NGC 4402, \citet{crowl08} measured a quenching time of $\sim$200 Myr in an area outside of the transition zone.  Given the transition zone width of 1.5 -- 2.5 kpc,  the stripping rate is between 7.5 pc Myr$^{-1}$ -- 12.5 pc Myr$^{-1}$.  

Alternatively, we can adopt a stripping rate in order to calculate a cloud survival time.  Another example of a similar-sized Virgo spiral undergoing active stripping, NGC 4330 \citep{abramson11}, has a UV gradient in the stripped outer disk indicating that 
it took $\sim$350 Myr for star formation to be quenched over $\sim$3.5 kpc in radius, giving a stripping rate of 10 pc Myr$^{-1}$.  Using this rate and the transition zone sizes gives cloud survival times of 50-150 Myr in NGC 4522, and 150-250 Myr in NGC 4402.  These survival times are similar to the ones calculated using stellar population quenching times in the two galaxies.  

Although we have calculated stripping rates and cloud survival times for a simple stripping scenario, in reality stripping may not be continuous or uniform.  It could vary rapidly due to ISM substructure, ICM substructure, or ICM motions.  However, we believe that the stripping rates and cloud lifetimes calculated in this section are plausible as time-averaged quantities.  

\subsection{Clouds Projected Against the Far Side of the Galaxy}
\label{farside}

Up to this point, we have focused our discussion on clouds projected against the near sides of the galaxies, since we can see almost all of the clouds projected against the near side that are in or above the disk plane.  In contrast, we can only see clouds projected against the far side if they are above the disk plane and are not projected against a larger dust complex.  Confusion results when a cloud located above the disk plane (on either the near or far side) is projected against the main dust lane or the large extraplanar dust complexes which cover a significant fraction of the far side.  Moreover, the extraplanar dust will block our view of decoupled clouds located behind it on the far side of the galaxy.  Clouds projected against the far side will only be detectable if they are projected beyond the dust truncation radius near the major axis, since the extraplanar material is preferentially located near the minor axis because of the ICM wind orientation.  Therefore, we do not have a complete census of clouds projected against the far side like we do with transition zone clouds projected against the near side.  Nonetheless, a significant number of the clouds we observe are projected against the far side - 11 of 13 in NGC 4522, and 4 of 20 in NGC 4402.  

Clouds projected against the far side are either located above the disk plane ($\gtrsim$1 kpc) on the far side, or well above the disk plane (1-3 kpc) on the near side.  The possible explanations for these extraplanar clouds depend on the density of the clouds.  If the clouds have typical GMC densities, they are too dense to strip directly by ram pressure.  Using typical values for Virgo ICM density ($\sim$10$^4$ cm$^{-3}$) at the locations of NGC 4402 and NGC 4522, and a galaxy-ICM velocity difference of 1500 km s$^{-1}$, ram pressure exceeds the gravitational restoring force at a radius of 3-5 kpc in the disks of these galaxies for surface densities of up to $\sim$10 M$_\odot$ pc$^{-2}$, which is roughly an order of magnitude less than GMC densities.  Therefore, if the extraplanar clouds have GMC densities, they must have formed in situ above the disk from stripped gas.  If they are less dense than GMCs, they may have been directly stripped from the disk, or they may have formed in situ from stripped gas.  With our current observations, we cannot tell which scenario is correct.  We note that simulations \citep{tonnesen10} suggest that dense gas can form from stripped lower density gas, so it is plausible that the extraplanar decoupled clouds have GMC densities and formed in situ.  

In physical terms, the near-side clouds do not differ significantly from the clouds projected against the far side.  Both sides of both galaxies have average measured cloud masses of a few $\times 10^4$, although there is the most variation among the near-side clouds in NGC 4402, which is the largest population of clouds.  The $a_V$ values do not differ significantly between the clouds on the near side and those projected against the far side in either galaxy.  Overall, the clouds projected against the far side do not have different elongations from near-side clouds.  In NGC 4402, 3 of the 4 clouds projected against the far side are elongated, and 4 of 10 are elongated in NGC 4522.  Since there are a few very large clouds in NGC 4402, cloud dimensions are best compared using a median size rather than an average.  The median width and length of all the clouds in both galaxies are 60 pc and 95 pc respectively, and 3 of 4 clouds projected against the far side in NGC 4402 exceed these values, along with 5 of 10 in NGC 4522.  In summary, there seems to be no fundamental difference between the clouds on the far side that we know must be above the disk plane and those on the near side, leading to the possibility that some of the near side clouds could be above the disk plane.    

We detect clouds projected against the far side just beyond the main dust lane.  In NGC 4402, clouds 1-4 are near a large dust complex visible in Figure \ref{4402atlas} on the eastern side of the disk.  In NGC 4522, clouds 3-13 are projected against the far side, and are located to the southwest of the main dust lane (Figure \ref{4522atlas}).  The majority of the clouds in NGC 4522 are projected against the far side, while only a few of the clouds in NGC 4402 are.  Below, we use our understanding of the ram pressure stripping process to give possible reasons why the galaxies have very different numbers of clouds projected against the near and far sides.  We have identified two possible causes: different stripping histories, and the effects of rotation and viewing angle.  

NGC 4522 has significantly more stripped extraplanar ISM (40\% of HI, and 10\% of H$\alpha$) than NGC 4402 (6\% of HI and 0.3\% of H$\alpha$), so it is no surprise that NGC 4522 also has more extraplanar decoupled clouds projected against the far side, regardless of whether the decoupled clouds were pushed directly from the disk or formed in situ from stripped lower-density material.  The larger amount of extraplanar material in NGC 4522 is indicative of more recent, strong ram pressure than NGC 4402, which has less extraplanar gas available to form clouds that might be projected against the far side.  The stronger pressure in NGC 4522 may have also destroyed most decoupled clouds in the disk plane, while the weaker pressure in NGC 4402  may have allowed more decoupled clouds to remain in the disk plane rather than being ablated or directly stripped.      

To further understand why the dust clouds are distributed differently in the galaxies, we consider the viewing geometry as it relates to the galaxy's motion towards or away from us, which provides a possible explanation for why most clouds are projected against the near side in NGC 4402 and the far side in NGC 4522.  For this discussion, it is helpful to divide the galaxies into upstream and downstream sides of the disk plane, and leading and trailing halves of the galaxy, with respect to the ICM wind.  This is illustrated for both galaxies in Figure \ref{geom}.  The leading half is affected more directly by the ICM wind, and the upstream side of the leading half is most directly exposed to ram pressure.  For both galaxies, the downstream direction turns out to be above the disk midplane, so we can detect clouds downstream of the disk in dust extinction.  

If most stripping and decoupling of dense clouds happens preferentially on the leading half, then by the time a stripped sector of the galaxy rotates around to the trailing half, many of the decoupled clouds may have already been destroyed, since the rotational period in the transition zone is $\sim$200-300 Myr and the transition zone clouds likely exist for $<$200 Myr (Section \ref{timescales}).  This could be part of the reason why we see so few decoupled clouds in the near-side transition zone of NGC 4522, which is nearer to the trailing half - they decoupled as the ISM was stripped $\sim$200 Myr ago and have since been destroyed by the ICM wind or pushed out of the disk plane.  NGC 4522's rotation means that ram pressure may have had longer to affect the clouds currently located on the near side.  In NGC 4402, the near-side transition zone is just upstream from the leading side, so we are able to see clouds that have decoupled more recently.       

\section{Summary \& Conclusions}
\label{conclusions}
NGC 4402 and NGC 4522 are Virgo Cluster spirals experiencing active ram pressure stripping that has caused radially truncated ISM distributions.  Using high-resolution HST B- and V-band images of dust extinction in the galaxies, we detect a number of decoupled clouds in each galaxy beyond the main ISM truncation radius, which have remained even after the surrounding ISM has been stripped away. We characterize the ``transition zone" as the area extending from 1 kpc below to 1 kpc above the disk plane, beyond the main ISM truncation radius, that is occupied by some of these decoupled clouds.    Some of the clouds are located in the transition zones, and some are projected against the far sides of the galaxies - they are too far above the disk planes to be considered in the transition zones and were likely stripped from the disk planes.  Some of the clouds have elongated morphologies, indicating that the transition zones are experiencing active ram pressure, and the outermost dust clouds in each galaxy have a PA consistent with the likely global ICM wind direction.    Beyond the transition zone, $\sim$1.5 kpc from the outer edge of the main dust lane, no dust extinction is detected.

We examine the size, mass, and spatial distribution of these decoupled clouds and propose that they correspond to GMCs.  We then estimate the percentage of the original ISM mass that remains in the transition zone, the fraction of pre-stripping star formation that is still present, and the lifetimes of decoupled clouds in the transition zone.   

\begin{enumerate}

\item In the transition zone, highly obscuring (0.25 $\leq$ $a_V$ $\leq$ 0.56), massive (probable M$_H$ = $10^4 - 10^5 M_\odot$), mostly 50-100 pc long decoupled clouds remain while the less-dense material around them has been stripped away.  The sizes and masses of the dust clouds are fairly consistent between the two galaxies, with the exception of some very large clouds in NGC 4402 ($\sim$500 pc long, $10^6$ M$_\odot$).  

\item We find a good correlation between dust cloud diameters and the HI + H$_2$ gas mass calculated via dust extinction, and the relationship in this work is consistent with previous studies of dust extinction in other nearby spiral galaxies.  The gas masses measured via the dust extinction technique are a factor of $\sim$10 lower for a given cloud diameter than those of GMCs measured using CO, probably due to the complicating factors of foreground light, cloud substructure, and resolution limitations.  We think our clouds are similar to GMCs and propose that their true masses are $\sim$10 times higher than the dust extinction technique indicates.  The factor of 10 correction is an upper limit on the correction - we do not think the clouds are denser than GMCs, but they could be as dense or less dense. 

\item  The upper limit of the amount of pre-stripping HI + H$_2$ ISM mass that remains in the transition zone is only $\sim$1-6\%, if we assume a smooth, exponential ISM distribution and use a factor of 10 correction to make our cloud masses consistent with those of GMCs with the same radius.  The remaining percentage is up to $\sim$11\% if the stripped region was originally underdense.  The survival fraction varies azimuthally within each transition zone, probably due to variations in the pre-stripping ISM substructure.  

\item Using ground-based H$\alpha$ images and assuming a smooth, exponential distribution of star formation prior to stripping, we find that the transition zones currently have star formation rates $\sim$2-3\% of their pre-stripping SFRs, which is comparable to the surviving HI + H$_2$ gas mass fraction.  Since the remaining H$\alpha$ and gas fractions are similar, there is no evidence that ram pressure is triggering star formation in the decoupled transition zone clouds.  

\item Based on star formation quenching times in or near the transition zones, the lifetimes of decoupled clouds in the transition zones may be up to 150-200 Myr.  Given the width of the transition zone, and assuming that the decoupled cloud survival time is equal to the stellar population quenching time, we estimate an average stripping rate of $\sim$3 - 30 pc Myr$^{-1}$ for NGC 4522 and 7.5 - 12.5 pc Myr$^{-1}$ for NGC 4402.

\item The mass of the decoupled clouds and the remaining star formation in the transition zone represent a small fraction of their previous values, and the lifetime of the decoupled clouds is relatively short.  We can infer that once the lower-density gas is stripped from part of the disk, star formation is effectively ended.

\item Both galaxies show signs that their pre-existing substructure is important in determining how far radially they are stripped at a given azimuthal location.  The effect of ISM substructure on a galaxy's stripping radius is not reflected in models that assume a smooth or uniform ISM.  

\item Many of the decoupled clouds are clearly extraplanar, and if they have GMC densities, they must have formed in situ from stripped gas.  If they are lower density, they may have formed in situ or been directly stripped from the disk plane.  Our current observations cannot distinguish between these scenarios.  
 
\end{enumerate}

\acknowledgments
We gratefully acknowledge Hugh Crowl and Tomer Tal for their help in obtaining and processing the data used in this paper.  We also gratefully acknowledge the referee for helpful comments that improved the paper.  Support for programs $\#$9773 and $\#$10528 was provided by NASA through a grant from the Space Telescope Science Institute, which is operated by the Association of Universities for Research in Astronomy, Inc., under NASA contract NAS 5-26555.


\bibliography{hstrefs}

\end{document}